\documentstyle[prd,aps]{revtex}

\input epsf

\begin{document}

\draft
\title{Post-Newtonian SPH calculations of binary neutron star 
coalescence. I. Method and first results.}
\author{Joshua A. Faber and Frederic A. Rasio}
\address{Department of Physics, Massachusetts Institute of Technology,
Cambridge, MA 02139}

\date{\today}

\maketitle

\begin{abstract}
We present the first results from our Post-Newtonian (PN) Smoothed
Particle Hydrodynamics (SPH) code, which has been used to study the
coalescence of binary neutron star (NS) systems.  The Lagrangian
particle-based code incorporates consistently all lowest-order (1PN)
relativistic effects, as well as gravitational radiation reaction, the
lowest-order dissipative term in general relativity.
We test our code on sequences of single NS models of varying
compactness, and we discuss
ways to make PN simulations more relevant to realistic NS
models.  We also present a PN SPH relaxation
procedure for constructing equilibrium models of
synchronized binaries, and we use these equilibrium models as initial
conditions for our dynamical calculations of binary coalescence.
Though unphysical, since tidal synchronization is not expected in
NS binaries, these initial conditions allow us to compare our PN work
with previous Newtonian results.

We compare calculations with and without 1PN effects, for NS with
stiff equations of state, modeled as polytropes with $\Gamma=3$.
We find that 1PN effects can play a major role in the coalescence,
accelerating the final inspiral and causing a significant misalignment
in the binary just prior to final merging.
In addition, the character of the gravitational wave signal is altered
dramatically, showing strong modulation of the exponentially decaying
waveform near the end of the merger. We also discuss briefly the implications
of our results for models of gamma-ray bursts at cosmological distances.
\end{abstract}


\section{Introduction and Motivation}

Coalescing compact 
binaries with neutron star (NS) or black hole (BH) components
provide the most promising sources of gravitational
waves for detection by the large laser interferometers 
currently under construction,
such as LIGO \cite{1}, VIRGO \cite{2}, 
GEO \cite{3,3a}, and TAMA \cite{4}.
In addition to providing a major new confirmation of
Einstein's theory of general relativity (GR), including the first direct
proof of the existence of black holes \cite{5,5a}, the detection of
gravitational 
waves from coalescing binaries at cosmological distances could provide 
accurate independent measurements of the Hubble constant
and mean density of the Universe \cite{6}.

Expected rates of NS binary coalescence in the Universe, 
as well as expected event rates in laser interferometers, have 
now been calculated by many groups (see \cite{7} for a recent review). 
Although there is some disparity 
between various published results,
the estimated rates are generally encouraging.
Finn \& Chernoff \cite{8} calculated that
an advanced LIGO could observe as many as 20 NS merger
events per year. This number corresponds to an assumed  Galactic NS
merger rate ${\cal R}\simeq10^{-6}\,{\rm yr}^{-1}$
derived from radio pulsar surveys \cite{9}. 
However, later revisions \cite{10}
increased this number to ${\cal R}\simeq3\times10^{-6}\,{\rm yr}^{-1}$,
using the latest galactic pulsar population model of 
Ref.~\cite{11}. This value is consistent with the  
upper limit of ${\cal R} 
\lesssim 10^{-4}\,{\rm yr}^{-1}$ derived (most recently) 
by Arzoumanian et al.\ \cite{12} on the basis of very general statistical 
considerations about radio pulsars, and by Kalogera \& Lorimer \cite{13}, who
studied constraints from supernova explosions in binaries.

Many calculations of gravitational wave emission from coalescing binaries 
have focused on the waveforms emitted during the last few thousand orbits, 
as the frequency sweeps upward from $\sim10\,$Hz to $\sim300\,$Hz.
The waveforms in this frequency range, where the sensitivity of ground-based 
interferometers is highest, can be calculated very accurately by 
performing post-Newtonian (hereafter PN)
expansions of the equations of 
motion for two point masses \cite{14}. However, at the end of the inspiral, 
when the binary separation becomes comparable 
to the stellar radii (and the frequency is $\gtrsim1\,$kHz), 
hydrodynamics becomes important and the character 
of the waveforms must change. 
Special purpose narrow-band detectors that can sweep up frequency in real 
time will be used to try to catch the last $\sim10$ cycles of the gravitational
waves during the final coalescence \cite{15}. These ``dual recycling''
techniques are being tested right now on the German-British interferometer
GEO 600 \cite{3a}. In this terminal phase of the coalescence,
when the two NS merge together into a single object, 
the waveforms contain information not just about the 
effects of GR, but also about the interior structure 
of a NS and the nuclear equation of state 
(EOS) at high density. 
Extracting this information from observed waveforms, 
however, requires detailed theoretical knowledge about all relevant
hydrodynamic processes. 
If the NS merger is followed by the formation 
of a BH, the corresponding gravitational radiation waveforms will also 
provide direct information on the dynamics of rotating core collapse
and the BH ``ringdown'' (see, e.g., Ref.~\cite{5}).

Many theoretical models of gamma-ray bursts (GRBs) 
have also relied on coalescing compact 
binaries to provide the energy of GRBs at
cosmological distances \cite{16}. 
The close spatial association of some GRB afterglows 
with faint galaxies at high redshifts is not inconsistent
with a compact binary origin, in spite of the large recoil
velocities acquired by compact binaries at birth \cite{16a}.
Currently the most popular models all assume that the coalescence leads
to the formation of a rapidly rotating Kerr BH surrounded by a torus
of debris. 
Energy can then be extracted either from the rotation of the BH or from
the material in the torus so that, with sufficient beaming, the
gamma-ray fluxes observed from even the most distant GRBs can be
explained \cite{17}. Here also, it is important to
understand the hydrodynamic processes taking place during the final 
coalescence before making assumptions about its outcome. For example,
contrary to widespread belief, it is not clear that the coalescence of
two $1.4\,M_\odot$ NS will form an object that must collapse to a BH,
and it is not certain either that a significant amount of 
matter will be ejected
during the merger and form an outer torus around the central object
(see Sec.~III below and Ref.~\cite{R2000a}).

The final hydrodynamic merger of two compact objects is driven by a combination
of relativistic and fluid effects. Even in Newtonian gravity,
an innermost stable circular orbit (ISCO) is imposed by
global hydrodynamic instabilities, which can drive 
a close binary system to rapid coalescence once the tidal interaction 
between the two stars becomes sufficiently strong.
The existence of these global instabilities 
for close binary equilibrium configurations containing a compressible fluid, 
and their particular importance for binary NS systems, 
was demonstrated for the first time by 
Rasio \& Shapiro (Ref.~\cite{RS13}, hereafter RS1--3 or collectively RS) 
using numerical hydrodynamic calculations.
These instabilities can also be studied using analytic methods.
The classical analytic work for close binaries containing an
incompressible fluid (e.g., Ref.~\cite{18}) was
extended to compressible fluids in the work of Lai, Rasio, \& Shapiro 
(Ref.~\cite{LRS15}, hereafter LRS1--5 or collectively LRS).
This analytic study confirmed the existence of dynamical 
instabilities for sufficiently close binaries.
Although these simplified analytic studies can give much physical
insight into difficult questions of global fluid instabilities, 
3D numerical calculations remain essential for establishing
the stability limits of close binaries accurately and for following 
the nonlinear evolution of unstable systems all the way to complete 
coalescence. 

A number of different groups have now performed such calculations, using
a variety of numerical methods and focusing on different aspects of the
 problem. Nakamura and collaborators (see \cite{19} and references therein)
were the first to perform 3D hydrodynamic calculations of binary 
NS coalescence,
using a traditional Eulerian finite-difference code. 
Instead, RS used the 
Lagrangian method SPH (Smoothed Particle Hydrodynamics). They focused
on determining the ISCO for initial binary models in strict
hydrostatic equilibrium and calculating the emission of gravitational waves
from the coalescence of unstable binaries. Many of the results of RS were
later independently confirmed by New \& Tohline \cite{20} and Swesty et
al. \cite{20a}, who used completely
different numerical methods but also focused on stability questions, and 
by Zhuge, Centrella, \& McMillan \cite{21,21a}, who also 
used SPH.  
Davies et al.\ \cite{22} and Ruffert et al.\ \cite{RJS,RRJ} have
incorporated a treatment of the nuclear physics in their hydrodynamic
calculations (done using SPH and PPM codes, respectively), motivated
by cosmological models of GRBs.
All these calculations were performed in Newtonian gravity, with
some of the more recent studies adding an approximate treatment of
energy and angular momentum dissipation through the gravitational 
radiation reaction \cite{23}.

Zhuge et al.\ \cite{21,21a} and Ruffert et al.\ \cite{RJS,RRJ} 
also explored in detail the dependence of
the gravitational wave signals on the initial NS spins.
Because the viscous timescales for material in the NS is much longer
than the dynamical timescale during inspiral, it is generally assumed
that NS binaries will be non-synchronized during mergers.  It is
generally found than non-synchronized binaries yield less mass loss
from the system, but very similar gravity wave signals, especially
during the merger itself when the gravity wave luminosity is highest
\cite{RJS}.  

All recent hydrodynamic calculations agree on
the basic qualitative picture that emerges for the final coalescence.
As the ISCO is approached, the secular orbital
decay driven by gravitational wave emission is dramatically accelerated
(see also LRS2, LRS3).
The two stars then plunge rapidly toward each other, and merge together 
into a single object in just a few rotation periods. In the corotating 
frame of the binary, the relative radial velocity of the two stars always 
remains very subsonic, so that the evolution is nearly adiabatic.
This is in sharp contrast to the case of a head-on collision between
two stars on a free-fall, radial orbit, where
shock heating is very important for the dynamics (see, e.g., RS1
and Ref.~\cite{23a}).
Here the stars are constantly being held back by a (slowly receding)
centrifugal barrier, and the merging, although dynamical, is much more gentle. 
After typically $1-2$ orbital periods following first contact,
 the innermost cores of the 
two stars have merged and the system resembles a single, very elongated ellipsoid.
At this point a secondary instability occurs: {\it mass shedding\/} 
sets in rather abruptly. Material (typically $\sim10$\% of the total mass) 
is ejected through the outer Lagrange
points of the effective potential and spirals out rapidly.
In the final stage, the inner spiral arms widen and merge together, 
forming a nearly axisymmetric torus around the inner, maximally rotating
dense core. 

In GR, strong-field gravity between the masses in
a binary system is alone sufficient to drive a close circular 
orbit unstable. 
In close NS binaries, GR effects combine nonlinearly
with Newtonian tidal effects so that the ISCO should be encountered
at larger binary separation and lower orbital frequency than 
predicted by Newtonian hydrodynamics alone, or GR alone for two point
masses. The combined effects
of relativity and hydrodynamics on the stability of close compact
binaries have only very recently begun to be studied,
using both analytic approximations
(basically, PN generalizations of LRS; see, e.g., \cite{24,LomRS}, 
as well as numerical
calculations in 3D incorporating simplified treatments of 
relativistic effects 
(e.g., \cite{25,25a}).
Several groups have been working on a fully relativistic
calculation of the final coalescence, combining the techniques of 
numerical relativity and numerical hydrodynamics in 3D \cite{26,26a}.  However
this work is still in its infancy, and only preliminary results
of test calculations have been reported so far.
It should be noted that 1PN calculations performed by Taniguchi and
collaborators \cite{26b} to study the location of the ISCO for corotating and
irrotational binaries find that the ISCO moves inwards as
post-Newtonian corrections are increased, due primarily to the effect
of 1PN potentials with momentum-based source terms present in the
system.  Similarly, 
Buonanno and Damour \cite{26c} find that the ISCO for point masses in
a binary under GR moves inwards with increasingly massive objects.

The middle ground between Newtonian and fully relativistic calculations
is the study of the hydrodynamics in PN gravity.  Formalisms exist
describing not only all lowest-order corrections (1PN) to Newtonian
gravity, but also the lowest-order (2.5PN) effects of the gravitational
radiation reaction \cite{27,BDS}. 
Such calculations have
been undertaken by Shibata, Oohara \& Nakamura \cite{28} using an Eulerian
grid-based code,
and more recently by Ayal et al.\ \cite{Ayal} using SPH.  
This paper is the first of a series in which we will present a 
comprehensive study of the hydrodynamics of compact binary coalescence
using a new PN version of
a parallel SPH code which we have been developing over the past two years.
This work will be the natural extension to PN gravity
of the original Newtonian study by RS.

PN calculations of NS binary coalescence are particularly
relevant for stiff NS EOS. Indeed, for most recent stiff EOS, the
compactness parameter for a typical $1.4\,M_\odot$ NS is in the
range $GM/Rc^2\simeq 0.1-0.2$, justifying a PN treatment.
After complete merger, an object
close to the maximum stable mass is formed, with $GM/Rc^2\simeq
0.3-0.5$, 
and relativistic effects become much more important. However, even then,
a PN treatment can remain qualitatively accurate if the final merged
configuration is stable to gravitational collapse on a dynamical
timescale (see the discussion in Sec.~\ref{sec:fate}).
Most recent theoretical calculations (e.g., the latest version of the
Argonne/Urbana EOS; see Ref.~\cite{29}) and a number of recent 
observations (e.g., of cooling NS; see Ref.~\cite{30}) provide 
strong support for a stiff NS EOS.  In this paper we represent NS with
stiff EOS by simple polytropes with an adiabatic exponent $\Gamma=3$
(i.e., the EOS is of the form $p_*=kr_*^\Gamma$, where $p_*$ is
the pressure, $r_*$ is the rest-mass density, and $k$ is a 
constant; see RS2 and LRS3, who obtain $\Gamma\simeq3$ for the
best polytropic fit to recent stiff NS EOS).

The most significant problem facing PN hydrodynamic simulations
is the requirement that all 1PN quantities be small
compared to unity.  Unfortunately, this precludes 
the use of realistic NS models.  Shibata, Oohara \&
Nakamura \cite{28} computed 1PN mergers of polytropes with $\Gamma=5/3$ and a 
compactness $GM/Rc^2=0.03$, leaving out the
effects of the gravitational radiation reaction.
Ayal et al.\ \cite{Ayal} performed calculations for polytropes with
$\Gamma=1.6$ or $\Gamma=2.6$ and compactness values in the range
$GM/Rc^2\simeq 0.02-0.04$,
including the effects of the gravitational radiation reaction.  
For comparison, a realistic NS of mass $M=1.4\,M_{\odot}$ and radius
$R=10\,{\rm km}$ has $GM/Rc^2=0.2$, i.e., about an order of
magnitude larger. Unfortunately, performing calculations
with artificially small values of $GM/Rc^2$ also has
the side effect of dramatically inhibiting the radiation reaction,
 which scales as $(GM/Rc^2)^{2.5}$.  

Our PN SPH code combines a new parallel version of the Newtonian 
SPH code used by RS with a treatment of PN gravity based on the 
formalism of Blanchet, Damour, and
Sch\"{a}fer (Ref.~\cite{BDS}, hereafter BDS).  
Our calculations include all 1PN effects, as well as a PN treatment of 
the gravitational radiation reaction. We have also developed
a relaxation technique by which accurate quasi-equilibrium configurations 
can be calculated for close binaries in PN gravity. These serve
as initial conditions for our hydrodynamic coalescence calculations.
In addition, we present in this paper a simple solution to
the problem of suppressed radiation reaction for models of NS with
unrealistically low values of $GM/Rc^2$. 

The outline of our paper is as follows. 
Section~II presents our numerical methods, including the description
of our new PN SPH code, a discussion of the advantages of 
using SPH for this work, and the steps taken to make our results as
realistic as possible. More details on the methods and
our treatment of the initial conditions are given in the appendices.
Section~III presents our initial
results, based on two large-scale simulations,
with and without 1PN effects. These are performed for synchronized
initial binaries containing two identical polytropes with $\Gamma=3$. 
In future papers, we will study systematically the dependence of these 
results on the NS EOS (by varying $GM/Rc^2$ and $\Gamma$, and using more 
realistic, tabulated EOS for nuclear matter at high density), the NS spins 
(allowing for nonsynchronized initial conditions), and the 
binary mass ratio. 
Motivation for this future work as well as a brief summary of our
present results are presented in Section~IV.

\section{Numerical Method}

\subsection{The SPH Code}

Smoothed Particle Hydrodynamics (SPH) is a Lagrangian method ideally
suited to calculations involving
self-gravitating fluids moving freely in 3D.
The key idea of SPH is to calculate pressure gradient forces by kernel
estimation, directly from the particle positions, rather than by finite
differencing on a grid (see, e.g., \cite{31}, for recent reviews on
the method).
SPH was introduced more than 20 years ago by 
Lucy, Monaghan, and collaborators \cite{32}, who used it to study dynamical 
fission instabilities in rapidly rotating stars. Since
then, a wide variety of astrophysical fluid dynamics problems have been
tackled using SPH (see Ref.~\cite{33} for an overview). 

Because of its Lagrangian nature,
SPH presents some clear advantages over more traditional 
grid-based methods for calculations of stellar interactions. Most
importantly, fluid advection, even for stars with a sharply defined surface
such as NS, is accomplished without difficulty in SPH, since the
particles simply follow their trajectories in the flow. In contrast, 
to track accurately the orbital motion of two stars across a large 3D grid
can be quite tricky, and the stellar surfaces then require a special
treatment (to avoid ``bleeding''). SPH is
also very computationally efficient, since it concentrates the
numerical elements (particles) where the fluid is at all times,
not wasting any resources on emty regions of space. For
this reason, with given computational resources, SPH provides higher
averaged spatial resolution than grid-based calculations, although
Godunov-type schemes such as PPM typically provide better 
resolution of shock fronts (this is certainly not a decisive advantage for
binary coalescence calculations, where no strong shocks ever develop).
SPH also makes it easy to track the hydrodynamic ejection of matter to 
large distances from the central dense regions. Sophisticated nested-grid 
algorithms are necessary to accomplish the same with grid-based
methods. 

Our simulations were performed using a modified version of
an SPH code that was originally
designed to perform 3D Newtonian calculations of stellar interactions  
(see Ref.~\cite{RS} and RS1).
Although the fluid description is completely Lagrangian, the gravitational 
field in our code (including PN terms) is calculated on a 3D grid using an
FFT-based Poisson solver.  
Our Poisson solver is based on the FFTW of Frigo \& Johnson \cite{FJ},
which features fully parallelized real-to-complex transforms.
Boundary conditions are handled by zero-padding all grids, which has
been found to produce accurate results and to be the most computationally 
efficient method (see also \cite{20a}).  Since we are
primarily interested in the gravitational wave emission, which originates
mainly from the inner dense regions of NS mergers, 
we fix our grid boundaries to be $\pm 4$ NS radii in
all directions from the center of mass.  Particles that fall outside
these boundaries are treated by including a simple monopole gravitational 
interaction with the matter on the grid.  Our code has been developed
on the SGI/Cray Origin 2000 parallel supercomputer at NCSA.
MPI (the Message Passing Interface) reduces 
the memory overhead of the code by splitting all large grids among 
the processors. 
All hydrodynamic loops over SPH particles and their neighbors 
have also been fully parallelized
using MPI, making our entire code easily portable to other parallel
supercomputers.  The parallelization provides nearly linear speedup
with increasing number of processors up to $\sim10$, with a
progressive degradation for larger numbers.

For more details on the Newtonian version of the code, and extensive
results from test calculations, see Ref.~\cite{LSRS}.
  
\subsection{The Blanchet, Damour, and Sch\"{a}fer PN Formalism with SPH}

To investigate the hydrodynamics of NS binary coalescence beyond the Newtonian 
regime, the equations of RS were modified to account for PN effects 
described by the formalism of BDS, 
converted into a Lagrangian, rather than Eulerian form.
The main equations and definitions of quantities appearing
in the BDS formalism are summarized briefly in Appendix~A.
The formalism is correct to first (1PN) order, with all new forces 
calculated from eight additional Poissson-type equations with compact support,
allowing for the computation of all 1PN terms using the same FFT-based
convolution algorithm as for the Newtonian Poisson solver. 
PN corrections to hydrodynamic quantities are calculated by the SPH 
method, i.e., by summations over particles.  Dissipation of energy 
and angular momentum by gravitational 
radiation reaction is included to lowest (2.5PN) order, requiring the
solution of one additional Poisson-type equation.

The key changes to the BDS formalism involve a conversion to
quantities based on SPH particle positions, rather than grid points.
In the discussion that follows, $a$ and $b$ refer to quantities
defined for particles, or particle neighbors, while $i$ and $j$ are
spatial indices.
The rest-mass density is calculated at each particle position as 
a weighted sum over the masses of neighboring particles,
\begin{equation}
r_{*}^{(a)}=\sum_b m_b W_{ab},
\label{SPHdensity}
\end{equation}
where $m_a$ is the rest mass of particle $a$, 
and the weights are given in terms of a smoothing kernel $W(\vec{r},h)$ by
\begin{equation}
W_{ab}=\frac{1}{2}[W(|\vec{r}_{(a)}-\vec{r}_{(b)}|, h_a)+
W(|\vec{r}_{(a)}-\vec{r}_{(b)}|, h_a)].  
\end{equation}
Here $h_a$ is the smoothing length for particle $a$,
which is updated after every iteration so as to keep the number of neighbors
as close as possible to a designated optimal value $N_N$.  The form of
the SPH kernel $W$ used in our calculations is the same standard
third-order spline used by RS (and most other current implementations of SPH).
This kernel is spherical and goes to zero for $r>2h$.

The total mass-energy of the system can be calculated as
\begin{equation}
M_g=\int d^3x~ r_*(1+\delta) = \sum_b m_b(1+\delta_b),
\end{equation}
where $\delta$ is a 1PN correction defined in Eq.~(\ref{pndelta}),
while the total rest mass is
\begin{equation}
M=\int d^3x~ r_* = \sum_b m_b.
\end{equation}

In our simple polytropic models of NS, the pressure is calculated 
from the local density as
\begin{equation}
p_*^{(a)}=k_a (r_{*}^{(a)})^{\Gamma}
\end{equation}
where $k_a$ is a function of the specific entropy of the particle, and
$\Gamma$ is the adiabatic exponent.  The standard Newtonian pressure
force is given by
\begin{equation}
F_i^{hydro}=-\frac{\partial_i p_*^{(a)}}{r_*^{(a)}}=   
-\sum_b m_b\left(\frac{p_*^{(a)}}{(r_*^{(a)})^2}+ 
\frac{p_*^{(b)}}{(r_*^{(b)})^2}\right)\nabla_i W_{ab}.
\end{equation}

In the absence of artificial viscosity (AV), entropy is conserved and
$k_a$ is constant throughout the simulation.  If AV is included,
the first law of thermodynamics can be expressed as
\begin{equation}
\frac{dk_a}{dt}=\frac{\Gamma-1}{2(r_*^{(a)})^{\Gamma-1}}\sum_b m_b 
\Pi_{(a,b)}(\vec{w}_{(a)}-\vec{w}_{(b)})\cdot\nabla_i W_{ab},
\label{dkdt}
\end{equation}
with a corresponding force on particles given by
\begin{equation}
F_i^{AV}=-\sum_b m_b \Pi_{(a,b)} \nabla_i W_{ab}.
\end{equation}
The expression for $\Pi_{(a,b)}$ depends on the particular choice of 
AV. We use the form proposed by Balsara \cite{34}, which gives
\begin{equation}
\Pi_{(a,b)}=\left(\frac{p_*^{(a)}}{(r_*^{(a)})^2}+ 
\frac{p_*^{(b)}}{(r_*^{(b)})^2}\right)
(-\alpha'\mu_{(a,b)}+\beta'\mu_{(a,b)}^2),
\end{equation}
where $\mu_{(a,b)}$ is a measure of the rate of convergence in the flow.
The exact definition of $\mu_{(a,b)}$ can be found in Eqs.~(15--19) of
Lombardi et al.\ \cite{LSRS}, who also show that the optimal choice of
the numerical coefficients is $\alpha'=\beta'=\Gamma/2$.
This form was shown to handle shocks properly and minimize the
amount of spurious mixing and numerical shear viscosity. 

For the PN pressure force (Eq.~\ref{fpress}), we now find
\begin{equation}
F_i^{press}=-(1+\frac{\alpha}{c^2})(F_i^{hydro}+F_i^{AV}) 
-\frac{1}{c^2}\frac{p_*}{r_*}\partial_i\alpha,
\end{equation}
where $\alpha/c^2 \ll 1$ is a PN correction.
In the calculation of $\partial_i\alpha=(2-3\Gamma)\partial_i
U_*-\frac{\Gamma}{2}\partial_i w^2$, we must take a
derivative of the local dynamic velocity-squared field, which we do by
SPH summations, i.e., we first write
\begin{equation}
\partial_i(w^2)=\frac{1}{r_*}(\partial_i(r_* w^2)-w^2\partial_i r_*),
\end{equation}
and we then calculate the derivative terms as
\begin{eqnarray}
\partial_i r_*^{(a)}&=&\sum_b m_b\partial_i W_{ab},\\
\partial_i(r_* w^2)^{(a)}&=&\sum_b m_b w_{(b)}^2 \partial_i W_{ab}.
\end{eqnarray}

The nine Poisson-type equations in the full PN formalism of BDS 
are all solved by the same FFT convolution method.  
All 3D grids used by the Poisson solver are distributed among
the processors in the z-direction.  Real-to-complex transforms are
computed using the {\sc rfftwnd\_mpi}
 package of the FFTW library \cite{FJ}.
The source terms of the Poisson equations that do not
contain density derivatives,
Eqs.~(\ref{poisustr},\ref{poispnux},\ref{poispnu2}), 
are laid down on the grid by a
cloud-in-cell method.  All integrals over the
density distribution are converted into sums over particles, e.g., 
\begin{equation}
U_*(x)=\int d^3x\frac{r_*(x')}{|x-x'|}\rightarrow\sum_b \frac{m_b}{|x-x_b|}.
\end{equation}
Source terms containing density
derivatives are calculated by finite differencing on the grid, rather
than by SPH-based derivatives at particle positions.  This has two
benefits. First, for integrals of the type 
\begin{equation}
\Phi=\int d^3 x~\partial_i r_* \ldots
\end{equation}
 we cannot convert directly from a volume integral to a sum
over discrete particle masses. Second, it guarantees that the volume
integral  of the source term vanishes in
Eqs.~(\ref{poispncx},\ref{poispnr}), as it should.  
Derivatives of the potentials are computed by finite differencing
on the grid, and then interpolated between grid points to assign values
at SPH particle positions.

The calculation of the quadrupole tensor and its
derivatives  are unchanged in a Lagrangian formulation.
When calculating the third derivative, however,
we found it advantageous to take the SPH expression for
the second derivative of the quadrupole tensor (RS1),
\begin{equation}
\ddot{Q}_{ij}=\sum_b m_b (v^{(b)}_i v^{(b)}_j+
x_i^{(b)}\partial_j U_{*}+x_j^{(b)}\partial_i U_{*}),
\end{equation}
and numerically differentiate once with respect to time.  The
resulting expression differs from the third derivative expression given 
in BDS by a term of $O\left(v^2/c^2\right)$, but all radiation
reaction terms into which it enters already contain factors of
$O\left(v^5/c^5\right)$. While only approximate, this method proved
more stable since it does not require the numerical evaluation
of several second derivatives on a grid.

For calculations in which we include the radiation reaction,
but ignore 1PN corrections, all terms
containing a factor of $1/c^2$ in Appendix~A can be ignored.  In this case
our equations reduce to those of the purely Newtonian
case, with two exceptions.  First, we include $F_i^{reac}$ 
(Eq.~\ref{freac}) in the SPH equations of motion, replacing
Eq.~(\ref{dotwi}) by
\begin{equation}
\dot{w_i}=-\frac{\partial_i p_*}{r_*}+\partial_i U_* +F^{reac}_{i}.
\label{wnewt}
\end{equation}

Second,
the relationship between the particle velocity $\vec{v}$ and momentum
$\vec{w}$ is given by 
\begin{equation}
v_i=w_i+\frac{4}{5}\frac{G}{c^5}Q^{[3]}_{ij} w_j.
\label{defvinewt}
\end{equation}
This has beeen shown \cite{RJS} to give the correct energy loss rate  as
predicted by the classical quadrupole formula,
\begin{equation}
\frac{dE}{dt}=\frac{1}{5}\frac{G}{c^5}<Q^{[3]}_{ij} Q^{[3]}_{ij}>.
\end{equation}
Ignoring 1PN terms reduces the number of Poisson equations to be solved per
iteration from nine to two.  The obvious advantage is a proper
handling of the dissipative PN effects, while leaving the 
hydrodynamic equations in a simple form that can be
directly compared to the Newtonian case.  In addition, because the
corrections are $O(v^5/c^5)$, the radiation reaction terms 
always remain small, even when 1PN corrections would be large.

We have performed a number of test calculations to establish the
accuracy of our treatment of PN effects in the SPH code. These include 
tests for
single rotating and nonrotating polytropes in PN gravity, which
we have compared to well-known analytic and semi-analytic results
\cite{ST}.  In particular, we have verified that our
code reproduces correctly the dynamical stability limit to radial
collapse for a single PN polytrope with $\Gamma=5/3$
(see Sec.~\ref{sec:single}).

\subsection{A Hybrid 1PN/2.5PN Post-Newtonian Formalism}

Throughout this paper, unless otherwise specified,
we use units in which Newton's gravitational
constant $G$, and the rest mass $M$ and radius $R$ of a single, spherical NS
are set equal to unity.
In Newtonian physics, this leads to a scale-free calculation (RS). When we
include PN effects, specifying the physical
mass and radius of the NS then sets the value of the speed of
light $c$, and the magnitude of all PN terms.  In our units, the
compactness ratio $GM/Rc^2$ of a NS is expressed simply 
as $1/c^2$.

The equations of BDS assume that all 1PN corrections are small.
As mentioned in Sec.~1,
this places a rather severe constraint on the allowed NS mass and
radius, since
\begin{equation}
\frac{1}{c^2}=0.14\left(\frac{M}{1.5\,M_{\odot}}\right)
\left(\frac{15\,{\rm km}}{R}\right).
\label{compact}
\end{equation}
If, for example, we estimate the
potential at the center of the star as $U_*/c^2\simeq 1.5/c^2=0.21$
 (Eq.~\ref{poisustr}), which is appropriate for $\Gamma=3$ models, we
find that our ``first-order'' correction term $\alpha/c^2$
(Eq.~\ref{pnalpha}), with $\Gamma=3$ and no internal motions, is
\begin{equation}
\frac{\alpha}{c^2}=(2-3\Gamma)\frac{U_*}{c^2}=-7\frac{U_*}{c^2}\simeq-1.5.
\end{equation}
This is clearly problematic since the derivation of the BDS formalism
assumes that $|\alpha|/c^2\ll 1$.  For a fixed radius of
$15\,$km and $\Gamma=3$, a NS mass $< 0.9\,M_{\odot}$, or
$1/c^2< 0.09$ is required to keep $|\alpha|< 1$. 
This problem is less severe for a lower value of $\Gamma$, since
the coefficient of $\alpha$ is then smaller.  For
$\Gamma=5/3$, we have $\alpha=-3 U_*$, but these configurations are
known to be unstable against gravitational collapse for compactness
parameters $1/c^2 \gtrsim 0.14$ (See, e.g., Ref.~\cite{ST}, and
Sec.~\ref{sec:single}).  These problems are the
reason why previous PN hydrodynamic simulations of NS binary
coalescence have used unrealistic NS models with low masses and large radii.
In practice, we find that we cannot calculate reliably NS mergers
including 1PN corrections, unless
$1/c^2 \lesssim 0.05$, or $c\gtrsim 4$.  
With such a small compactness parameter, radiation reaction effects
would then be
suppressed by a factor $\sim 2^5=32$.

Recognizing that the 1PN and 2.5PN terms describe essentially
independent phenomena,
and that the proper form for energy and angular momentum loss holds even 
if 1PN corrections are ignored, we adopt a hybrid scheme.  Specifically,
in this paper, we
set $c=4.47\equiv c_{1PN}$ for all 1PN corrections, which is unphysically
large, but we use a physically realistic value of $c=2.5\equiv c_{2.5PN}$
for the 2.5PN corrections,
corresponding, for example, to a NS mass $M=1.5\,M_{\odot}$ with radius
$R=13.9\,{\rm km}$.
We feel that this hybrid formulation provides a reasonable trade-off
between physical reality
and the limitations of the 1PN approximation.  

Note that this method should better extrapolate toward physical
reality, compared with unrealistically undercompact NS models.
If $0+2.5$PN simulations are interpreted as taking the limit
$c_{1PN}\rightarrow\infty$ for the 1PN corrections, we see that by reducing
the compactness in both the $0+2.5$PN and $0+1+2.5$PN cases, the value
of $c_{2.5PN}$ is fixed at an unphysical value while $c_{1PN}$ is
varied, which can never truly extrapolate to the physical case.  By
setting $c_{2.5PN}$ to a realistic physical value while varying
$c_{1PN}$, we may be able to extrapolate our results toward a correct
physical limit.
However, a disadvantage of this approach is that it does not allow for
direct quantitative comparison with full GR simulations of binary NS
coalescence. In these simulations, which essentially handle
corrections to all orders simultaneously, separation into various PN
orders has no meaning.

\subsection{Initial Conditions}

In addition to its normal use for dynamical calculations, our SPH 
code can also be used to construct hydrostatic equilibrium 
configurations in 3D, which provide accurate initial conditions
for binary coalescence calculations.
This is done by adding artificial friction terms to the fluid equations
of motion and forcing the system to relax to a minimum-energy state
under appropriate constraints (RS).
The great advantage of using SPH itself for setting up 
equilibrium solutions is that the
dynamical stability of these  solutions can then be tested immediately by
using them as initial conditions for dynamical SPH calculations.
Very accurate 3D equilibrium
solutions can be constructed using such relaxation techniques, with the
virial theorem satisfied to better than 1 part in $10^3$ and
excellent agreement found with known quasi-analytic solutions 
in both Newtonian (LRS1, LRS4, RS2) and PN gravity
\cite{LomRS}. 
The careful construction of accurate quasi-equilibrium initial
conditions is a distinguishing feature of both our previous
Newtonian calculations (RS)  and our new PN calculations of
binary coalescence. In contrast, most other studies have used
very crude initial conditions, placing two spherical stars
in a close binary orbit, and, for calculations that went beyond
Newtonian gravity, adding the inward radial velocity for the
inspiral of two point masses. As demonstrated in Sec.~\ref{sec:inspiral}, 
this leads to a significantly slower inspiral rate. Moreover, spurious
fluid motions are created as the stars respond dynamically to
the sudden appearance of the strong tidal force. These 
can in turn corrupt the gravitational radiation waveforms.
Spurious velocities have additional effects
in the full 1PN case, where spurious
motions enter repeatedly into the evolution equations, by
propagating through the 1PN quantities $\alpha$, $\beta$,
and $\delta$ (Eqs.~\ref{pnalpha},\ref{pnbeta},\ref{pndelta}).
A specific cause of worry is the influence of velocities adding to
$\delta$, which affects not only the self-gravity of the
stars, but also their mutual gravitational attraction.

We have developed a method, described in detail in Appendix~B, 
that allows for more realistic
initial conditions for PN synchronized binaries.  
It reduces dramatically the initial oscillations
around equilibrium when the dynamical calculation is started.
Since the method varies rather significantly between the full
$1PN+2.5PN$ formalism and the Newtonian with radiation reaction
formalism, we handle the two cases separately.  For the Newtonian
case, radiation reaction plays no role in the relaxation, entering
only into the initial values of the velocity and momentum, $\vec{v}$
and $\vec{w}$ (Eq.~\ref{defvi}), upon the start of the dynamical run.
The 1PN case is considerably more complicated, requiring the
construction of static single star models, which are then input into a
PN binary relaxation scheme.

\section{Results}

We have performed two large-scale hydrodynamic simulations of NS binary
coalescence, with and without the 1PN correction terms of BDS.  Both
simulations included radiation reaction throughout the entire run,
treated in the formalism of Appendix A.
Hereafter, we refer to these two runs as the
Newtonian (N) and post-Newtonian (PN) runs, noting that the N simulation did
include 2.5PN effects. 

For both runs, we used 50,000 particles per NS (total of $10^5$), 
with a $\Gamma=3$
polytropic EOS.  The two NS are identical. 
Synchronized rotation was assumed in the initial condition.
The optimal number of neighbors for each SPH particle was set to 100.
Shock heating, which plays a completely negligible role in the case
studied here, was ignored and therefore the SPH AV was turned off.  
All Poisson equations were solved on grids of size $256^3$, including 
the added space necessary 
for zero-padding.  For the 1PN run, we used a compactness parameter
$1/c_{1PN}^2=0.05$ (see Sec.~IIC).  
In both runs, we used $c_{2.5PN}=2.5$ in calculating
radiation reaction terms.  The N run required a total of 600 CPU hours
and the PN 
run 1200 hours on the NCSA Origin2000, including the relaxation phase. 
Particle plots illustrating qualitatively
the evolution of the system are shown in Fig.~\ref{fig:xyplotn} (N~run) 
and Fig.~\ref{fig:xyplotp} (PN run).

\subsection{Dynamical Instability and the Inspiral Process}
\label{sec:inspiral}
It was shown by RS and LRS that equilibrium configurations for close
binary NS become dynamically unstable when the separation $r$ is less than
a critical value.  For Newtonian, synchronized, equal-mass binaries 
with $\Gamma=3$, the ISCO is at $r=2.95\,R$.  Purely Newtonian
calculations for binaries starting from equilibrium configurations with a
separation larger than this value will show no evolution in the
system.  Binaries starting from a smaller separation, though, are
dynamically unstable, and coalesce within a few orbital periods, even
without the energy and angular momentum loss due to radiation reaction
(RS,\cite{20,20a}). 

In simulations with radiation reaction included, coalescence will
always be the end result.  The limiting factor on how large to make 
the initial separation is the computing time required for the binary orbit
to slowly spiral inward.  Ideally, one should make sure that the stars
are in quasi-equilibrium when the orbit approaches the ISCO
and the inspiral timescale undergoes a shift from the
slow radiation-reaction timescale to the much faster dynamical timescale.

Since the effective gravitational attraction between two stars is
increased by PN effects, we expect the ISCO to
move outwards when 1PN corrections are included.  This was demonstrated
by Lombardi, Rasio, \& Shapiro \cite{LSRS}, who used the same energy 
variational method as LRS to
find equilibrium configurations for binary NS models including 1PN
corrections. Taking into account these results,
we used an initial separation of $r_0=3.1\,R$ for our
N~simulation, and $r=4.0\,R$ for the PN simulation.  
As a consequence, there is an ambiguity in the relative time between 
the two runs, which
we resolve by adjusting the initial time of the N run such that the
maximum gravity wave luminosity occurs at the same time in both the N
and PN runs.  This was found to require shifting the time in the 
N~run backwards
so that it starts at $t=-13$, while the PN run starts at $t=0$. 

In Fig.~\ref{fig:sepplots}, we show the evolution of the
center-of-mass binary separation during the initial inspiral phase
for our N and PN runs.   Fig.~\ref{fig:sepn} shows the
inspiral phase of the N~run, as well as the
inspiral tracks predicted by the classical quadrupole formula for two
point masses, and by the methods of LRS3 for two corotating spheres
and two ellipsoids.  We note that the results of LRS3 predict for
extended objects
a significantly more rapid inspiral rate, which is confirmed by the
numerical run.  In addition, we note that the approach of the ISCO
is clearly visible in the plots, where the inspiral rate
switches from the slow radiation-reaction-driven orbital decay
to the faster dynamical infall.  This appears to happen at $r\simeq 2.7\,R$ 
in the N case, in good agreement with previous results.

Comparing the PN run to the N run, we see that the
stability limit must lie at a considerably larger separation.  This
agrees with the results of Lombardi et al.~\cite{LSRS}, who
find that PN corrections not only move the ISCO
outward, but also flatten out the equilibrium binary energy
curve $E(r)$ near the stability limit (where $E(r)$ reaches a minimum).  
Following the arguments of LRS3, we
conclude that unstable inspiral begins when the differential change in
binary energy as a function of separation becomes smaller than the
energy loss rate to gravitational radiation.  The condition for
unstable inspiral, expressed as 
\begin{equation}
\frac{dE}{dr}\ll \frac{dE_{GW}}{dt}\left(\frac{dr}{dt}\right)^{-1},
\end{equation}
is then encountered further outside the ISCO (as determined
for binaries in strict equilibrium), since PN
corrections decrease the left-hand side.  This effect can also be seen
in the results of Ayal et al.~\cite{Ayal} by careful examination
of their Fig.~5a.  Even though the binary separation 
in their PN run has a large initial oscillation, caused by the use of
non-equilibrium initial conditions, it still
converges at a much more rapid rate than in their
corresponding Newtonian model. 

Even though the effective stability limits of N and PN
binaries differ by a large amount, their actual inspiral velocities are
essentially the same from the moment of first contact, at a
separation of $r\simeq 2.5\,R$, until the merger of the NS
cores.  The only significant difference is the break in the inspiral
velocity for the N run at $t\simeq20$, which occurs as the cores
start to come into direct contact with each other.  The lack of this
feature in the PN run will be explained in Sec.~\ref{sec:coalesce}.

\subsection{Coalescence}
\label{sec:coalesce}

In Fig.~\ref{fig:rhoplots}, 
we show the time evolution of the maximum density in both runs.
The maximum density is at the center of either star initially, but
it shifts eventually to the center of the merger remnant. 
The initial oscillations with a period of $T\simeq 2-3$ correspond to the
fundamental radial pulsations of the polytropes, and represent the
errors resulting from small departures from strict equilibrium in the 
initial conditions.  We see that
$\delta\rho/\rho\simeq 0.01$ and $0.05$, respectively, for the
N and PN runs, which provides a measure of the numerical accuracy of
the initial conditions.

As the binary system contracts to separations of $r\lesssim 2.7\,R$,
we see a rather sudden and rapid decrease in the maximum density found
at the core of each star, corresponding closely with the
moment of first contact of the two stars, after which the cores 
get tidally stretched. For the PN run, this
follows a gradual increase in the average density maximum, which is
caused by the contraction of each NS in response to the growing
gravitational potential of its companion, rather than a pure tidal
effect.  This effect, which seems to result primarily from the 
weakening of the pressure force in Eq.~(\ref{fpress}) as $\alpha$
becomes more negative in response to the growing gravitational
potential (Eq.~\ref{pnalpha}), was also seen by Ayal et al.\ in one
of their runs (Ref.~\cite{Ayal}, see their Fig.~6, run P3).
When the center of mass
separation reaches a value of $r\simeq 2.0\,R$ the maximum density
stops decreasing, turning around and increasing sharply as the cores
come into direct contact and merge.

In Fig.~\ref{fig:gwplots}, we show the gravity wave
signatures of both runs. The waveforms in the two polarizations 
of gravitational radiation are calculated for an observer at a distance
$d$ 
along the rotation axis of the system in the quadrupole approximation,
\begin{eqnarray} 
c^4 (d\,h_+)&=&\ddot{Q}_{xx}-\ddot{Q}_{yy} \label{hplus}
\\
c^4 (d\,h_{\times})&=&2\ddot{Q}_{xy} \label{hcross}.
\end{eqnarray}
In Fig.~\ref{fig:gwlum} we show the corresponding gravity wave
luminosity of the system, given by
\begin{equation}
c^5 L_{GW}=\frac{1}{5}\left\langle Q^{[3]}_{ij}
Q^{[3]}_{ij}\right\rangle. \label{gwlum}
\end{equation}
We see that, as the inner NS cores merge, the 
gravity wave luminosity peaks for both runs, with the characteristic
frequency of the waves increasing like (twice) the rotation frequency of
the system.  This frequency increase is more rapid in the PN
case, since the inspiral is faster.

After $t\simeq30$, the evolution of the N binary is
rather straightforward.  A triaxial object is formed at the center of 
the system, with spiral outflows emanating from the outer parts of each
star. The spiral arms remain coherent for several windings before slowly
dissipating, and finally leaving a low-density halo of material 
in the region $r/R\simeq2-15$. 
During this time, the central triaxial object acts as the predominant
source for the gravity waves as it spins down, leading to a
characteristic damped oscillatory signature, at a luminosity
approximately $1/30$ that of the peak.  The rise in central density
from the initial value at $t=0$ to the final value at $t=80$ is
consistent with what is expected from the mass-radius relation
for a Newtonian polytrope with $\Gamma=3$.

This simple picture, which is familiar from many previous
Newtonian simulations, is
seen to break down when 1PN effects are taken into account.
As is clear from Fig.~\ref{fig:xyplotp} ($t=12$), just
prior to the final coalescence, 1PN effects
cause the long axis of each star to rotate forward relative to 
the binary axis, so that the inner part of each
star leads the center of mass in the orbital rotation. This
{\em dynamical tidal lag\/} is expected from the rapidly changing
tidal forces during the final inspiral phase (LRS5). It is not
to be confused with the tidal lag produced by viscous dissipation 
in nonsynchronized binaries (see, e.g., Ref.~\cite{34a}).
The dynamical tidal lag angle can be estimated analytically for a
Newtonian binary whose orbit decays slowly by gravitational
wave emission. Using Eq.~(9.21) of LRS5, we estimate a lag angle 
$\alpha_t\simeq 0.01$ for $1/c^2=0.16$
and $r\simeq2R$. This is in agreement with the very small
lag angle observed in our N run (barely visible at $t=12$ in
Fig.~~\ref{fig:xyplotn}). In contrast, from our PN run, we find 
$\alpha_t\simeq 0.14$, indicating that the more rapid inspiral
can dramatically increase this effect.

As the PN merger proceeds, material from
the leading edge of each star wraps around the other, so that the
cores simply slide past each other instead of striking more nearly 
head-on as in the N case.  
As this happens around $t\simeq 25$, 
the maximum density drops slightly,
and the gravity wave luminosity rises again, reaching a second peak at
$t\simeq35$, with a maximum luminosity $L_2=0.65\,L_1$ compared to the first
peak of luminosity $L_1$.  A cursory examination of Fig.~\ref{fig:xyplotp}
reveals a highly asymmetric, triaxial configuration near this time.  
The subsequent oscillations of the two cores in their sliding motion
against each other
damp out rather quickly, and the central object becomes more nearly
axisymmetric while the maximum density rises again. 
A third peak of maximum
luminosity $L_3=0.15\,L_1$ is clearly visible near $t\simeq51$, as is
another very slight drop in the central density at that time, 
and a fourth, much smaller luminosity peak occurs at $t\simeq72$.  

To better understand this oscillation of the merger, and the 
corresponding modulation of the gravitational radiation waveforms,
we show in Fig.~\ref{fig:mom} a comparison between the gravity wave
luminosity and the ratio of the principal moments of
inertia of the central object in the PN run.  
As can be seen clearly, the two quantities are strongly correlated.  
If we ignore the details of the internal motion of the fluid, 
it may be tempting to model the late-time behavior of the remnant 
in terms of a simple quadrupole 
($l=2$ f-mode) oscillation of a rapidly and uniformly rotating single
star. Adopting an average value for the angular velocity 
of the central object, $\bar{\Omega}^2=0.4$, 
and using Eq.~(3.30)
of LRS5 for the frequency of the quadrupole oscillation of a
compressible Maclaurin spheroid, we obtain a frequency 
$\sigma=0.38$, which gives us a modulation period $T_{\rm mod}=16.6$, 
very close to what we observe in Figs.~6 and~7.

The occurrence of a second peak in the gravity wave luminosity 
can also be seen in the PN calculations presented in Ref.~\cite{Ayal}
for polytropes with $\Gamma=2.6$, but the second peak appears 
considerably less pronounced for $\Gamma=2.6$ than for $\Gamma=3$.  
This may simply result from the higher central concentration of
objects with lower values of $\Gamma$, which decreases the emission
of gravitational radiation for a quadrupole deformation of given
amplitude.
Grid-based Newtonian calculations by Ruffert et al.\ \cite{RJS} for
nonsynchronized 
binaries with a different EOS also show a second peak in the gravity-wave
luminosity.
For Newtonian systems with  $\Gamma\lesssim2.2$, 
the merger remnant evolves quickly to axisymmetry and the
emission of gravitational radiation stops abruptly after
the first peak (cf.\ RS1 and RS2).

\subsection{The Final Merger Product}\label{sec:finalstate}

In Fig.~\ref{fig:finalcontour}, we show density contours of the central 
merger remnant in both the
equatorial and vertical planes.  For the N run, the remnant is shown at
$t=80$, which is at the end of the calculation.  For the PN run, we 
show the remnant at $t=55$, which corresponds to the third gravity 
wave luminosity peak,
and at $t=80$, the end of the simulation and close to a gravity wave 
luminosity {\em minimum\/}. 
Axes for the contour plots are aligned with the principal 
axes of the remnant.  A summary of values for the principal axes 
and moments of inertia for
the three configurations is presented in Table~\ref{table:final}.

We see that the final remnant in the PN case is larger and
more centrally condensed than in the N case, with a higher
degree of flattening in the vertical direction.
This is in part because in the PN case less mass and angular momentum
is extracted from the central region and deposited in the halo.  Figures
\ref{fig:jm3n} and \ref{fig:jm3p} show the evolution of the angular
momentum of the various components in both runs.
In the N case, most of the angular momentum lost by the remnant has
gone into the halo.  In the PN case, about equal amounts of angular
momentum are lost to the halo and to the gravity waves.

Nevertheless, the axis ratio $a_2/a_1$ in the equatorial plane is
approximately the same for the N run at $t=80$ and for the PN run at
$t=55$ and at $t=80$, indicating a reasonably constant shape for the
outermost region. Further comparison between the N and
PN remnants, however, shows that their interior structures
are remarkably different.  In the PN remnant,
the isodensity surfaces do not maintain a consistent
orientation or shape as we move from the center to the equator of
the remnant,
indicating that the structure of the remnant is much more complex than
that of a self-similar ellipsoid.
Gravity-wave luminosity peaks are seen to occur when the inner and 
outer contours are aligned, leading to a larger net quadrupole moment
(this is nearly the case at $t=55$ in Fig.~\ref{fig:finalcontour}).  
Minima occur when the orientations lie at right angles, as can be seen
near $t=80$ for the PN run in Fig.~\ref{fig:finalcontour}.   

In Fig.~\ref{fig:finalstate}, we show the radial mass and rotational
velocity profiles of the remnant.  Horizontal cuts through the matter
indicate that the rotation is cylindrical, with rotational velocity a
function only of the distance from the rotation axis, independent of
height relative to the equatorial plane (the same type of rotation
profile has been obtained from strictly Newtonian calculations; see RS1).  
Neither case gives a rigid
rotation law.  The angular velocity of the N remnant shows a
slight increase for $r>1.1\,R$, whereas the PN run shows a decreasing
angular velocity at the same point.  Thus, both exhibit signs of
differential rotation, but in opposite directions.  
We find that the centrifugal acceleration and
gravitational acceleration become equal at the outer edge of the
remnant for both cases, at $r\simeq 1.6\,R$ and $r\simeq 1.85\,R$
for the N and PN runs, respectively.  This is in good agreement with
the morphology of the remnants  
seen in Fig.~\ref{fig:finalcontour}, where a noticeable cusp-like
deformation is visible in the outermost density contours near  the
equator in the vertical plane.  We conclude that
in both runs, the final remnant is {\em maximally and differentially\/}
rotating.
	
The rest mass of the N remnant at $t=80$ is $M_r=1.73\,M$, while
that of the PN remnant $M_r=1.90\,M$.  The remaining mass, $0.27\,M$
for the N run and $0.10\,M$ for the PN run, has been shed during the
coalescence, forming the spiral arms seen in the middle panels of
Figs.~\ref{fig:xyplotn} and~\ref{fig:xyplotp}.  These spiral arms
later merge to form a halo of matter around the central remnant.
With a crude linear extrapolation from a halo mass of 
$M_h=0.27\,M$ for the N run, with
$1/c^2_{1PN}=0$, and $M_h=0.10\,M$ for the PN, run with
$1/c^2_{1PN}=0.05$, we might expect that, for physically reasonable NS
with $1/c^2\sim 0.15-0.20$, the vast majority
of the mass will remain in the central remnant.
However, this result may be crucially dependent on
our choice of initial spins and the EOS, and it is limited by the
restrictions we have placed on the magnitude of the 1PN corrections.  
It should also be noted that fully GR calculations of the coalescence of
NS with a $\Gamma=2$ EOS suggest that significant mass loss occurs even for
extremely compact NS \cite{26a}. 

\subsection{The Final Fate of the Remnant}\label{sec:fate}

By their very nature our calculations cannot address directly the question 
of whether the NS merger remnant will collapse 
to form a BH. Indeed the parameters of our PN run were chosen
so that all 1PN quantities remain small throughout the evolution, which,
for $\Gamma\gg 4/3$, guarantees stability. 
This can be verified directly by checking, for example, that
the mass distribution in the final merger remnant remains
everywhere well outside the corresponding horizon radius
(see Fig.~\ref{fig:finalstate}). However, given some of the general
properties of the merger remnant as determined by our calculations,
we can ask whether an object with similar properties, but with
a more realistic EOS and higher compactness, would still remain 
stable to collapse in full GR. 
For the coalescence of two $1.4\,M_\odot$ NS
with realistic stiff EOS, it is by no means
certain that the core of the final merged configuration will collapse
on a dynamical timescale to form a BH (see Refs.~\cite{R2000a,35}
 for recent discussions).  

The final fate of a NS binary merger in full GR depends not only
on the NS EOS and compactness, but also on the rotational state
of the merger remnant.
It has been suggested, for example, that the Kerr
parameter $a_r\equiv J_r/M_{gr}^2$ 
of the remnant may exceed unity for extremely 
stiff EOS \cite{25}. This does not appear to be the case, at least
for our choice of EOS. In Fig.~\ref{fig:jm2},
we show the evolution of the Kerr parameter throughout the entire
coalescence, including only particles for which the rest-mass density
satisfies $r_*>0.005$.  This cut includes essentially 
all matter in the initial stages, and effectively cuts out 
particles in the spiral outflow once the coalescence begins, as well as
those remaining in the outer halo at the end.  
We see that $a_r$ is very near unity just prior to the final merger, 
but, in contrast to what has been assumed in some previous studies
\cite{35}, it decreases significantly
during the final coalescence.  
The decrease occurs mainly during periods of maximum gravity-wave
luminosity, as angular momentum is radiated away, and during 
the mass-shedding phase after $t\simeq 20$, since 
angular momentum is transferred from the core to the outside spiral
outflow. By the end of the PN run, $a_r$ has decreased to $\simeq0.7$,
well below unity, and certainly not large enough to prevent collapse.
The final value of the Kerr parameter for the PN run, $a_r=0.70$, is
considerably greater than that of the N run, $a_r=0.47$.  The difference
is attributable to the greater mass ejected in the N run,
which carries off a significant fraction of the angular momentum of the
system (see Figs.~\ref{fig:jm3n} and \ref{fig:jm3p}).  

Quite apart from considerations of the Kerr parameter,
the rapidly rotating core may be dynamically stable. Indeed, most 
stiff NS EOS (including the
recent ``AU'' and ``UU'' EOS of Ref.~\cite{36}) allow stable,
maximally rotating NS with baryonic masses exceeding
$3\,M_\odot$ \cite{37}, i.e., well above the mass
of the final merger core (which is $1.9\,M\simeq 2.85\,M_\odot$
for $M=1.5\,M_\odot$ in our PN calculation; see Fig.~{\ref{fig:finalstate}).
Differential rotation (not taken into account in the
calculations of Ref.~\cite{37}) 
can further increase this maximum stable 
mass very significantly (see \cite{35}).
For slowly rotating stars, the same EOS give
maximum stable baryonic masses in the range $2.5-3\,M_\odot$,
implying that the core would probably (but not certainly)
collapse to a BH in the absence of rotational support. 

If the final merger remnant is being stabilized against collapse by 
rotation, one must then
consider ways in which it may subsequently loose angular momentum.
Further reduction of the angular momentum of the core by gravitational
radiation
or dynamical instabilities cannot occur, since, at the end of the 
dynamical coalescence,
the core is, by definition, dynamically stable and nearly axisymmetric
(i.e., no longer radiating gravity waves; see Figs.~6 and~7). 
The development of a secular bar-mode instability (a quadrupole mode
growing unstably on the viscous dissipation timescale; see LRS1 and LRS4)
has been discussed as a way of reducing the angular momentum of a
rapidly rotating compact object
\cite{37a}.  However, 
this cannot occur either for a binary merger remnant because,
if the remnant were rotating fast enough to be secularly unstable,
it would still be triaxial (Recall, for example, that the point of
bifurcation of the classical Maclaurin spheroid sequence into the
Jacobi ellipsoid sequence coincides with the onset of secular
instability for Maclaurin spheroids; see, e.g., \cite{ST} and LRS1).
Note that other processes, such as 
electromagnetic radiation or neutrino emission,
which may also lead to angular
momentum losses, take place on timescales much longer than the dynamical
timescale (see, e.g., Ref.~\cite{38}, where it is shown that
neutrino emission is probably negligible). These processes are
therefore decoupled from the hydrodynamics of the coalescence.
Unfortunately their
study is plagued by many fundamental uncertainties in the microphysics.

\section{Summary and Directions for Future Work} 

Using a Lagrangian, SPH-based adaptation of the BDS PN formalism for
hydrodynamics, we have calculated the merger of a coalescing
NS binary including 1PN and gravitational radiation reaction effects.
We have also developed a method for computing accurate, quasi-equilibrium
initial data for coalescing binaries in PN gravity, improving upon
previous calculations that used nonequilibrium initial conditions
containing unperturbed, spherical stars.

We have confirmed that PN corrections to gravity cause the binary
inspiral to become dynamical at larger binary separation compared to
what is predicted in the Newtonian limit.
In calculations using Newtonian gravity, but including the effects
of the gravitational radiation reaction, we have found that the
inspiral rate just prior to merging agrees well with the predictions
of semi-analytic models using compressible ellipsoids as trial functions
in an energy variational method (LRS).

Using a hybrid formalism where radiation reaction is treated realistically
but 1PN effects are reduced in amplitude so as to remain numerically
tractable, we have compared the hydrodynamic coalescence of binary NS
systems in Newtonian and PN gravity.  We find that 1PN effects lead to  
important qualitative differences in the hydrodynamic behavior and in the
gravitational radiation waveforms and luminosities.
In Newtonian gravity, the merger of two equal-mass, $\Gamma=3$
polytropic NS produces a single peak in the gravity-wave luminosity,
followed by an exponentially decaying signal.  In PN gravity, we see a
strong quadrupole oscillation of the remnant immediately after
coalescence, which leads to several additional peaks in the gravity-wave
luminosity.  
In both Newtonian and PN gravity, the final merger remnant is 
found to be maximally rotating and nearly axisymmetric.  
Even for realistic NS EOS and in full GR, this
configuration is expected to be stable against gravitational collapse 
to a BH on a dynamical timescale. The amount of mass ejected into
an outer halo by the rotational instability developing during the final
merger decreases substantially when 1PN effects are included, and we
suggest that, for realistic NS models, essentially no mass might be ejected,
so that the total baryonic mass of the system remains entirely in the 
central remnant (though this result is hardly a certainty).

Our study is naturally beginning with PN calculations for equal-mass, corotating
binaries with a simple polytropic EOS. This allows us
to compare our results directly to previous Newtonian calculations
performed with the same set of assumptions (RS,\cite{20,20a,RRJ}).
The dependence of our results on the NS EOS
will be studied in future papers by varying the adiabatic exponent $\Gamma$ (in the
range $\Gamma\simeq2-4$ applicable to NS; see, e.g., LRS3) and by
performing additional runs with more realistic tabulated NS EOS. In
particular, we will consider the Lattimer-Swesty EOS \cite{39}.
This EOS includes high-temperature effects (which can be significant
in the outermost, low-density regions of some NS mergers) and has also
been employed in several previous Newtonian studies \cite{23},
to which we want to compare our results. Even with the lowest
available value of the nuclear compressibility ($K=180\,$MeV), 
the Lattimer-Swesty EOS is
relatively stiff (effective $\Gamma\simeq2.5$ for a $1.4\,M_\odot$
NS). The latest microscopic NS EOS, constrained by nucleon scattering
data and the binding of light nuclei, and incorporating three-body
forces, are even stiffer (effective $\Gamma\gtrsim3$; see, e.g., 
Ref.~\cite{40} for a summary, and Ref.~\cite{29} for the latest version).
We will use several of these recent EOS, in tabulated form, to
perform additional, more realistic calculations.
More schematic EOS based on exotic states of matter, such as 
pion condensates or strange quark matter, can be much softer
($\Gamma\lesssim2$ and maximum stable masses not much above $1.4\,M_\odot$).
We will not consider such soft EOS in our calculations, since
they render the PN approximation invalid. Note that
several observations in progress may have already ruled them out 
(e.g., from the large measured mass of the NS in Vela X-1; \cite{41}).

Using our PN SPH code we will also study the dependence of the 
hydrodynamics and gravitational wave emission on the binary mass
ratio $q$. Neutron star masses derived from observations
of binary radio pulsars are all consistent with a
remarkably narrow underlying Gaussian mass distribution with
$M_g=1.35\pm0.04\,M_\odot$ \cite{42}.
The largest observed departure from $q=1$
in any known binary pulsar with
likely NS companion is currently $q=1.386/1.442=0.96$
for the Hulse-Taylor pulsar PSR B1913+16 \cite{43}.
Although the equal-mass case is
clearly important, one should not conclude from these observations
that it is unnecessary to consider coalescing NS binaries with
unequal-mass components. Indeed, it cannot be excluded that other 
binary NS systems (that may not be observable as binary pulsars) 
could contain stars with significantly different masses. 
Moreover, Newtonian calculations of binary NS coalescence have shown
that even very small departures from $q=1$ can drastically affect
the hydrodynamic evolution (RS2,\cite{21a}).

In future papers we will also perform PN SPH calculations for binary 
NS systems that are initially {\it nonsynchronized\/}. This is likely to be the case
for real systems, since the tidal synchronization time in close NS
binaries is probably always longer than the orbital decay
time \cite{44}.
The methods of LRS can be used to construct approximate,
quasi-equilibrium initial conditions for nonsynchronized coalescing 
binaries. For binaries that are far from synchronized,
the final coalescence involves
some new, complex hydrodynamic processes, and significant differences in the 
gravitational wave emission compared to the synchronized case, with an additional
dependence of the gravitational radiation waveforms on the stellar spins
\cite{21a,45}.
Moreover, the final fate of the merger may also be very different for
initially nonsynchronized binaries, since the merger remnant may no longer be
maximally rotating \cite{R2000a,RJS}.

\acknowledgements
This work was supported in part by NSF Grant AST-9618116 and NASA ATP 
Grant NAG5-8460. 
J.A.F.\ acknowledges support from a Karl Taylor Compton
Graduate Fellowship at MIT.
F.A.R.\ was supported in part by an Alfred P.\ Sloan Research Fellowship.
This work was also supported by the National Computational Science Alliance under grant
AST980014N and utilized the NCSA SGI/CRAY Origin2000.

\appendix
\section{The BDS $1+2.5~PN$ Formalism}
In the original Eulerian, PN formalism of Blanchet,
Damour, and Sch\"{a}fer \cite{BDS}, the key variables appearing in the
hydrodynamic evolution equations are PN variants of the
standard Newtonian quantities.  Specifically, the coordinate
rest-mass density $r_*$ and momentum per unit rest-mass $w_i$
are given in terms of the proper rest-mass density $\rho$ and the 4-velocity 
$u^{\mu}$ by 
\begin{eqnarray}
r_*&=&\sqrt{g} u^0\rho \label{defrstr}\\
w_i&=&\left(1+\frac{h}{c^2}\right)cu_i, \label{defwi}
\end{eqnarray}
where $h$ is the specific enthalpy of the fluid.  Assuming hereafter a
polytropic equation of state, i.e. one for which the pressure is given
by
\begin{equation}
p_*(r_*)=kr_*^\Gamma \label{defpstr},
\end{equation}
it is found that the specific enthalpy is given by
\begin{equation}
h=k\frac{\Gamma}{\Gamma-1}{r_*^{\Gamma-1}}= 
\frac{\Gamma}{\Gamma-1}\frac{p_*}{r_*}. \label{defh}
\end{equation}
It should be noted that $p_*$ is not the Newtonian pressure, but rather a 1PN
variant of it.

The BDS formalism requires the solution of nine Poisson equations, one
for the Newtonian gravitational potential $U_*$, seven for 1PN
corrections, and a final one to handle the gravitational radiation
reaction.

The equation for the gravitational potential is
\begin{equation}
\nabla^2 U_*=-4\pi r_*. \label{poisustr}
\end{equation}
Note that with this sign convention, the
gravitational potential is a positive quantity.
The 1PN correction potentials are given by
\begin{eqnarray}
\nabla^2 U_i&=&-4\pi r_*w_i \label{poispnux}\\
\nabla^2 C_i&=&-4\pi x^i\partial_s(r_* w_s) \label{poispncx}\\
\nabla^2 U_2&=&-4\pi r_* \delta. \label{poispnu2}
\end{eqnarray}
Note that the summation in Eq.~(\ref{poispnux}) runs over $i=x,y,z$, thus
$U_2\neq U_y$ in Eq.~(\ref{poispnu2}).
Using these, we define the quantity
\begin{equation}
A_i=4U_i+\frac{1}{2} C_i-\frac{1}{2} x^i\partial_s U_s. \label{defpnax}
\end{equation}

It is important to note that the volume integral of the source term of
Eq.~(\ref{poispncx}) vanishes, assuming that the origin is at the 
center of mass and momentum of the system, and thus it contains no
monopole term.  In Eq.~(\ref{poispnu2}), the quantity $\delta$ in the
source term is one of three quantities which are assumed to be of
order $O(\frac{1}{c^2})$.  They are, assuming the equation of state
Eq.~(\ref{defpstr}), and with $w^2=\delta^{ij}w_i w_j$,
\begin{eqnarray}
\alpha&=&2U_*-\Gamma(\frac{1}{2} w^2+3U_*)\label{pnalpha}\\ 
\beta&=&\frac{1}{2} w^2+\frac{\Gamma}{\Gamma-1}\frac{p_*}{r_*}+3U_*
\label{pnbeta}\\
\delta&=&\frac{3}{2}w^2+\frac{3\Gamma-2}{\Gamma-1}\frac{p_*}{r_*}-U_*.
\label{pndelta}
\end{eqnarray}
The third derivative of the symmetric, trace-free (STF) quadrupole
tensor, $Q^{[3]}_{ij}$ is calculated from
\begin{eqnarray}
P_{ij}&=&2\int d^3x r_*[3w_i \partial_j U_*-2w_i\frac{\partial_jp_*}
{r_*}+x^iw_s\partial_{sj}U_*-x^i\partial_{sj}U_s] \label{defp3ij}\\
Q^{[3]}_{ij}&=&\frac{1}{2} P_{ij}+\frac{1}{2} P_{ji}
-\frac{1}{3}\delta_{ij}P_{ss}, \label{defq3ij}
\end{eqnarray}
and is used in the source term for the radiation reaction potential
$U_5$, of order $O(\frac{1}{c^5})$.  This is calculated from the final
Poisson equation,
\begin{eqnarray}
U_5&=&\frac{2}{5}G\left(R-Q^{[3]}_{ij}x^i\partial_j{r_*}\right)
\label{defpnu5} \\
\nabla^2 R&=&-4\pi Q^{[3]}_{ij}x^i\partial_j r_*. \label{poispnr}
\end{eqnarray}
Since we are dealing with the trace-free quadrupole tensor, it is easy
to show that the volume integral of the source
term of Eq.~(\ref{poispnr}) also vanishes, for any mass distribution.

Forces are defined by
\begin{eqnarray}
F_i^{press}&=&-\left(1+\frac{\alpha}{c^2}\right)\frac{\partial_i
p_*}{r_*}-\frac{1}{c^2}\frac{p_*}{r_*}\partial_i \alpha
\label{fpress}\\
F_i^{1PN}&=&\left(1+\frac{\delta}{c^2}\right)\partial_i U_*
+\frac{1}{c^2}\partial_i U_2 -\frac{1}{c^2}w_s\partial_i A_s
\label{fpn}\\
F_i^{reac}&=&\frac{1}{c^5}\partial_i U_5. \label{freac}
\end{eqnarray}

Finally, the evolution system, in Eulerian form, is given by
\begin{eqnarray}
\partial_t r_*&=&\partial_i(r_* v^i) \label{dotrstr}\\
\partial_t w_i&=&-v^s\partial_s w_i+F_i^{press}+F_i^{1PN}+F_i^{reac},
\label{dotwi}
\end{eqnarray}
where the particle velocities $v^i$ are related to the specific
coordinate momentum $w_i$ by
\begin{equation}
v^i=\left(1-\frac{\beta}{c^2}\right)w_i+\frac{1}{c^2}A_i+
\frac{4}{5}\frac{G}{c^5}w_s Q^{[3]}_{is}. \label{defvi}
\end{equation}
The quantities $\vec{v}$ and $\vec{w}$ will be referred to simply 
as the velocity and momentum vectors, respectively (see \cite{RJS}).

In the SPH method, the evolution equations must be expressed in a
Lagrangian form, given simply by
\begin{eqnarray}
\dot{x}^i&=&v^i \label{dotx2}\\
\dot{w}_i&=&F_i^{press}+F_i^{1PN}+F_i^{reac}. \label{dotwi2}
\end{eqnarray}

In BDS, there also appear evolution equations for the entropy
and the pressure.  The former, converted into Lagrangian form, states
that entropy is a conserved quantity, and is handled in our code by
the choice of AV, as in Eq.~(\ref{dkdt}).
The latter is not necessary here
since we calculate the pressure directly from the density at each time step.

Since the parameters $\alpha$ and $ \beta$, defined by 
Eqs.~(\ref{pnalpha},\ref{pnbeta}) become rather large
for NS with $1/c^2\sim 0.05$, we make some small adjustments to
Eqs.~(\ref{fpress},\ref{defvi}).  We note that for a adiabatic exponent
$\Gamma>\frac{2}{3}$, $\alpha$ is everywhere negative.  
To ensure that the pressure force always
acts in the proper direction, we make a substitution
in Eq.~(\ref{fpress}),
\begin{equation}
-\left(1+\frac{\alpha}{c^2}\right)\frac{\partial_i p_*}{r_*} 
\rightarrow -\left(1-\frac{\alpha}{c^2}\right)^{-1} 
\frac{\partial_i p_*}{r_*}.
\end{equation}
This new form is entirely equivalent to the one it replaces to 1PN order.
Similarly, $\beta$ is everywhere positive, so we make the following
substitution in 
Eq.~(\ref{defvi}),
\begin{equation}
\left(1-\frac{\beta}{c^2}\right) w_i 
\rightarrow \left(1+\frac{\beta}{c^2}\right)^{-1} w_i.
\end{equation}

\section{Relaxation Methods}

\subsection{PN Case}\label{sec:single}

Constructing hydrostatic equilibrium initial conditions in PN gravity 
is a much more difficult problem
than in Newtonian gravity, primarily because of the complex
relationship between the particle velocity and momentum.
We get around this problem by implementing a
multistage approach to the construction of relaxed configurations.

First, we construct a series of hydrostatic equilibrium models for
single $\Gamma=3$ polytropes with increasing
values of $1/c^2$, to gauge the
effects of the PN corrections on the structure of the stars.
Specifically, we construct relaxed models with compactness parameters of
$1/c^2=0.01$ to $0.07$, in steps of $0.01$. 

In the relaxation procedure, 
spurious velocities arising from configurations adjusting toward
equilibrium are ignored as sources for the force equations.  
Thus particles move during the relaxation, but the force exerted on
each particle is that of a static mass configuration.
We thus solve all Poisson equations assuming
$\vec{w}=0$, which eliminates Eqs.~(\ref{poispnux},\ref{poispncx}).  
In addition, the velocity terms in the definition of the 1PN quantities
$\alpha$, $\beta$, and $\delta$ are removed from Eqs.\
(\ref{pnalpha},\ref{pnbeta},\ref{pndelta}).
This greatly simplifies the equations giving us the
set
\begin{eqnarray}
\label{singlestar}
\nabla^2 U_*&=&-4\pi r_* \\
\nabla^2 U_2&=&-4\pi r_* \delta \\
\alpha&=&(2-3\Gamma)U_*\\ 
\beta&=&\frac{\Gamma}{\Gamma-1}\frac{p_*}{r_*}+3U_*\\
\delta&=&\frac{3\Gamma-2}{\Gamma-1}\frac{p_*}{r_*}-U_*\\
F_i^{press}&=&-\left(1+\frac{\alpha}{c^2}\right)\frac{\partial_i
p_*}{r_*}-\frac{1}{c^2}\frac{p_*}{r_*}\partial_i \alpha\\
F_i^{1PN}&=&\left(1+\frac{\delta}{c^2}\right)\partial_i U_*
+\frac{1}{c^2}\partial_i U_2\\
v^i&=&\left(1-\frac{\beta}{c^2}\right)w_i\\
\dot{x}^i&=&v^i \label{dotx}\\
\dot{w}_i&=&F_i^{press}+F_i^{1PN}-\frac{w_i}{t_{relax}},
\label{singlestar2}
\end{eqnarray}
where $t_{relax}$ is the relaxation time.

To construct our first model, with $1/c^2=0.01$
we start from a Newtonian
$\Gamma=3$ polytrope and let it relax to an equilibrium configuration.
Then, using the maximum particle radius 
$R_{max}$, we  adjust the radial position, smoothing length, and specific
entropy of all particles according to 
\begin{eqnarray}
\vec{r}&\rightarrow&\frac{\vec{r}}{R_{max}} \\
h_m &\rightarrow& \frac{h_m}{R_{max}} \\
k_m &\rightarrow& k_m R_{max}^{4-3\Gamma}. 
\end{eqnarray}
Velocities are set to zero at the end of this rescaling.
This new configuration is relaxed again, and the process is repeated 
until convergence is achieved.  For the $\Gamma=3$ models, we rescaled
every $t=5.0$, with a relaxation time $t_{relax}=1.0$,
The final profile is used as the initial test configuration of the
next model, which is then relaxed iteratively as described above.

In addition to $\Gamma=3$ models, we also computed a sequence of single
PN polytropes with $\Gamma=5/3$, 
and tested their stability.  PN effects shouldmake the star unstable to
gravitational collapse when $1/c^2\gtrsim0.141$ for
$\Gamma=5/3$ \cite{ST}.  We tested $1/c^2$ values in
steps of $1/c^2=0.02$, until we reached 0.10, at which point we
halved the step size until we reached $1/c^2=0.13$.  To make the
relaxation overdamped, we reduced the rescaling time to $t=2.0$,
with $t_{relax}=1.0$.
It was found
that $1/c^2=0.13$ is always unstable, collapsing inward
uncontrollably, no matter how short the
rescaling time.  This agrees well with the theoretical
prediction when we account for the magnitude of the 1PN
corrections we deal with, and the approximations made in the analytic
treatment.  In Fig.~\ref{fig:plotak}, we show the time evolution of
the  specific entropy
$k$ for both sequences, taken as a
ratio with the Newtonian value of the specific entropy derived from
the Lane-Emden equation.  We see a gradual increase of $k$ as the
compactness is increased, in both sequences, until we get
to $1/c^2=0.12$ for $\Gamma=5/3$, for which $k$ is 
$50\%$ larger than the corresponding value for $1/c^2=0.11$.  

Parameters for the single star sequences are shown in Table
\ref{table:stuffvsc}.  Radial profiles of the density, as well as all
important 1PN quantities are shown in Figs.~\ref{fig:single} and 
\ref{fig:single2}. 
We see in the $\Gamma=5/3$ case that increasing the compactness
increases the central concentration of the model, which can be seen in
the factor of 2 increase in central density.   For compactnesses near
the stability limit, we see that $\alpha$, $\beta$, and $\delta$
are all of order unity.  A different behavior is seen in the $\Gamma=3$
case, for which the internal structure of the star remains almost
unchanged as the 1PN order parameters get large.  We see that
$\alpha$ and $\beta$ both get relatively large for more compact models, but
$\delta$ is rather small, since the potential and pressure terms
cancel each other to some extent.

 A comparison of the mass profile for the
$\Gamma=3$ polytrope with $1/c^2=0.05$, the model used in the PN dynamical
simulation, to a direct Runge-Kutta integration of the 1PN structure equations
in spherical symmetry
is shown in Fig.~\ref{fig:masrk}.  We see excellent agreement, except
at the outer edge of the star, where surface effects alter the SPH mass
profile slightly.  This results from a layer of particles developing
at the surface of the stars, with a slight density decrement
immediately within, but involves only a very small fraction of the
total mass of the system.
Since our method restricts particle positions to
$r/R<1$, we see that the density falls to zero slightly outside this point
because of the finite size of the SPH smoothing kernel.

Once these single star configurations were complete, the resulting
stars were placed in duplicate in a binary configuration, which was
assumed to be in a state of synchronized rotation, i.e.,  the velocity
of every SPH particle is given as a function of position by
\begin{equation}
\vec{v_0}=\vec{\Omega} \times \vec{r}.
\end{equation}

The main difficulty in relaxing PN configurations is in
the interplay between $\vec{v}$ and $\vec{w}$, which not only
differ in magnitude but also in direction.  Thus, one or the other can
be relaxed in the corotating frame, but not both. Here $\vec{v}$ was
assumed to be zero in the corotating frame  for a relaxed
configuration, satisfying the equation above.  

We created a method to calculate $\vec{w}_0(\vec{v}_0)$, which is
not invertible in closed form.  As can be seen from Eq.~(\ref{defvi}),
the relationship between particle velocity and momentum is a function
of several potentials at the particle position, through the term containing
$A_i$.  Since $A_i$ is itself a function of $\vec{w}$ (see
Eq.~\ref{defpnax}), and vice versa,
we need to solve consistently for both.
It was found to be best to use an
iterative procedure, which alternately solves for $\vec{w}$ and then
uses these trial values in the source terms of the relevant Poisson
equations. 

In the initial step, using known values of $\vec{v}_0$, we first
approximate $\vec{w}_0$ by the equations 
\begin{eqnarray}
\beta_{test}&=&\frac{1}{c^2}\left(3U_* +\frac{\Gamma}{\Gamma-1}
\frac{p_*}{r_*}\right) \\
\vec{w}_0&=&\vec{v}_0\left(1+\frac{\beta_{test}}{c^2}+
\frac{v_0^2}{2c^2}\left[1+\frac{\beta_{test}}{c^2}\right]^2\right).
\end{eqnarray}
The computed value of $\vec{w}$ 
enters into the source terms of both $U_i$ and $C_i$
(Eqs.~\ref{poispnux},\ref{poispncx}).  Using
these two potentials, we calculate $A_i$ and $\beta$
(Eqs.~\ref{defpnax},\ref{pnbeta}), and
recalculate a new approximation to $\vec{w}_0$, denoted  
$\vec{w}_{new}$, from the previous one, $\vec{w}_{old}$,  
by an iterative method, using only $\frac{1}{3}$ of the
correction to avoid overshooting, thus
\begin{equation}
\vec{w}_{new}=\frac{2}{3}\vec{w}_{old}+\frac{1}{3}\left(1+\frac{\beta}{c^2}
\right)\left(\vec{v}_0-\frac{\vec{A}}{c^2}\right). 
\end{equation}
It was found that, for the models we tested, about ten iterations would give
convergence to within 1 part in $10^3$ to the correct value of $\vec{v}$ when
compared to the value of $\vec{v}(\vec{w}_{new})$
calculated by Eq.~(\ref{defvi}).
For every timestep afterwards, 
we followed the same iteration procedure, and about six iterations were
found to produce the same convergence to the proper values.

Once convergence to an 
 acceptable solution was found, forces were calculated, and
$\dot{v}$ was estimated by finite differencing,
\begin{equation}
\dot{v}_{force}=\frac{\vec{v}(\vec{w}(t+dt))-\vec{v}(\vec{w}(t))}{dt}.
\end{equation}
We relax the binary models at fixed center-of-mass separation $r$, in the
corotating frame, adjusting
$\Omega$ such that the inward force of gravity is balanced exactly by
the centrifugal force.  At every time step, we calculate
\begin{equation}
\Omega=\sqrt{\frac{F^1_{in}+F^{2}_{in}}{2r}},
\label{omeg}
\end{equation}
where $F_{in}$ refers to the net inward force on each component of the
binary.  
Particle velocities are advanced according to
\begin{equation}
\dot{v}=\dot{v}_{force}-\frac{v}{t_{relax}}+\Omega^2 r.
\end{equation}
After every time step, the two stars were adjusted slightly to
maintain a center of mass separation at the desired value.

\subsection{Newtonian Case}

In the regime where the dynamical timescale of the neutron stars
is much smaller than the characteristic timescale for gravitational 
radiation, we expect the stars to evolve through a series of
quasi-equilibrium configurations.  If synchronized rotation is
assumed, these equilibrium configurations can be constructed by adding
a centrifugal force and drag term to the acceleration equation, giving
us
\begin{equation}
\dot{v}_i=F_i^{hydro}-\nabla_i(\Phi+\Phi_{rot})-\frac{v_i}{t_{relax}},
\end{equation}
where the centrifugal potential is given by
\begin{equation}
\Phi_{rot}=\frac{1}{2}\Omega^2(x^2+y^2).
\end{equation}
The relaxation timescale, $t_{relax}$ is set initially to 1.0, close to
the value required for critical damping of oscillations (RS1).
For the purposes of relaxation, AV and the radiation
back-reaction, which are both time-asymmetric, are ignored.  In
addition, during the relaxation, we ignore the distinction between
velocity and momentum vectors in Eq.~(\ref{defvinewt}), taking 
 $\vec{v}=\vec{w}$.
The rate of rotation is calculated as in the PN case by 
Eq.~(\ref{omeg}).
Once the binary has relaxed to a suitable initial configuration, it is
set in motion, and we commence the dynamical run.
Initial velocities are given by
\begin{equation}
\vec{w}_x=-\Omega y,~\vec{w}_y=\Omega x,
\end{equation}
and $\vec{v}$ is calculated from 
$\vec{w}$  by Eq.~(\ref{defvi}).  In the point mass
limit, this would reduce to Eq.~(35) of Ruffert, Janka, and
Sch\"{a}fer \cite{RJS}, who use
\begin{equation}
v_r=-\frac{16}{5}\frac{M^3}{r^3}
\end{equation}
as their initial condition.

\newpage
\begin{figure}
\centering \leavevmode \epsfysize=7in \epsfbox{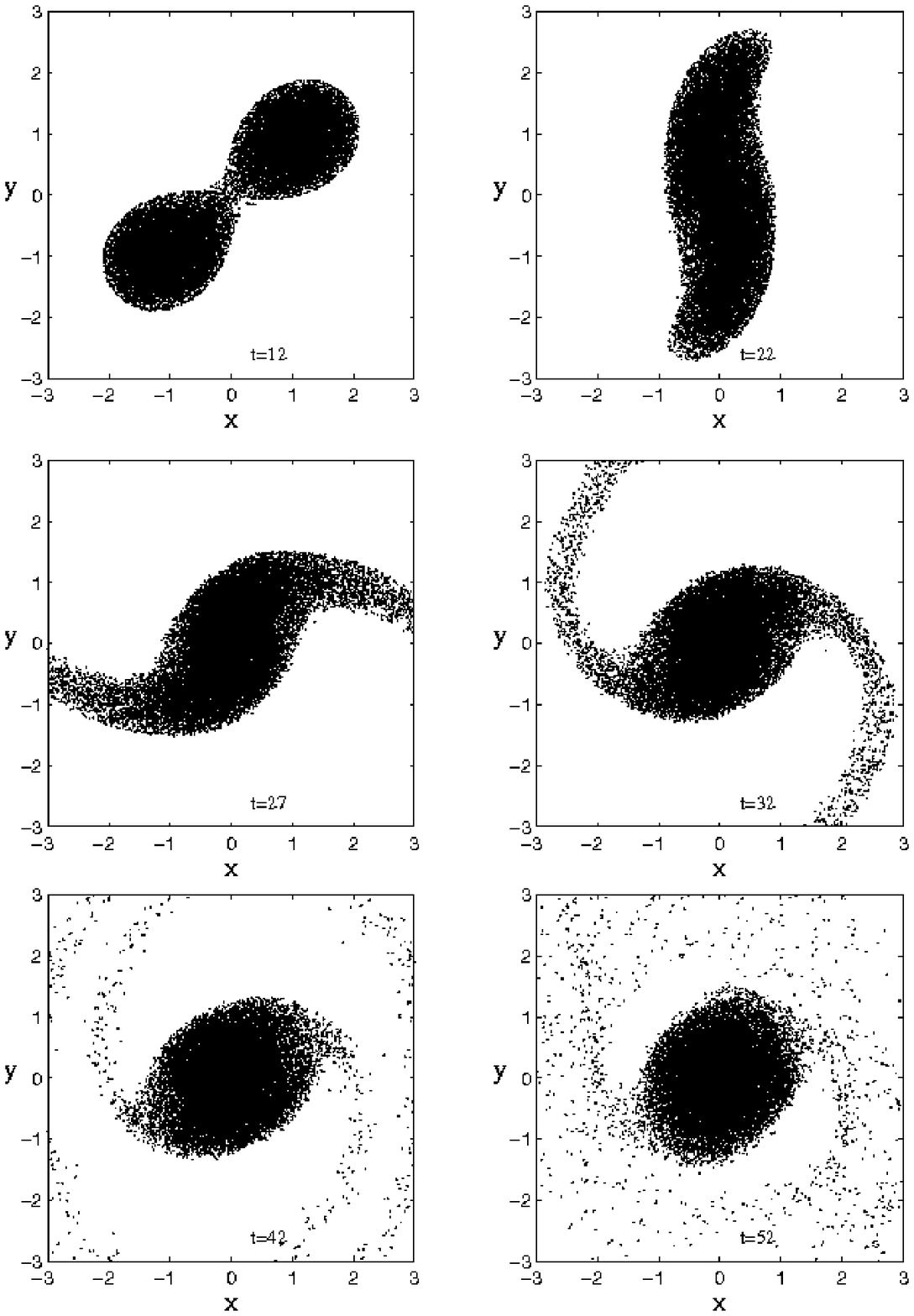}
\caption{Evolution of the system in the N run.  Projections of a
random subset of $20\%$ of all SPH particles onto the orbital (x-y)
plane are shown at various times.  The orbital motion is
counter-clockwise.  Units are such that $G=M=R=1$, where $M$ and $R$
are the mass and radius of a single, spherical NS.  Note that the
development of a mass-shedding instability after $t\simeq 25$, and the
rapid contraction of the remnant toward an axisymmetric state at late times.}
\label{fig:xyplotn}  
\end{figure}
\newpage
\begin{figure}
\centering \leavevmode \epsfysize=7in \epsfbox{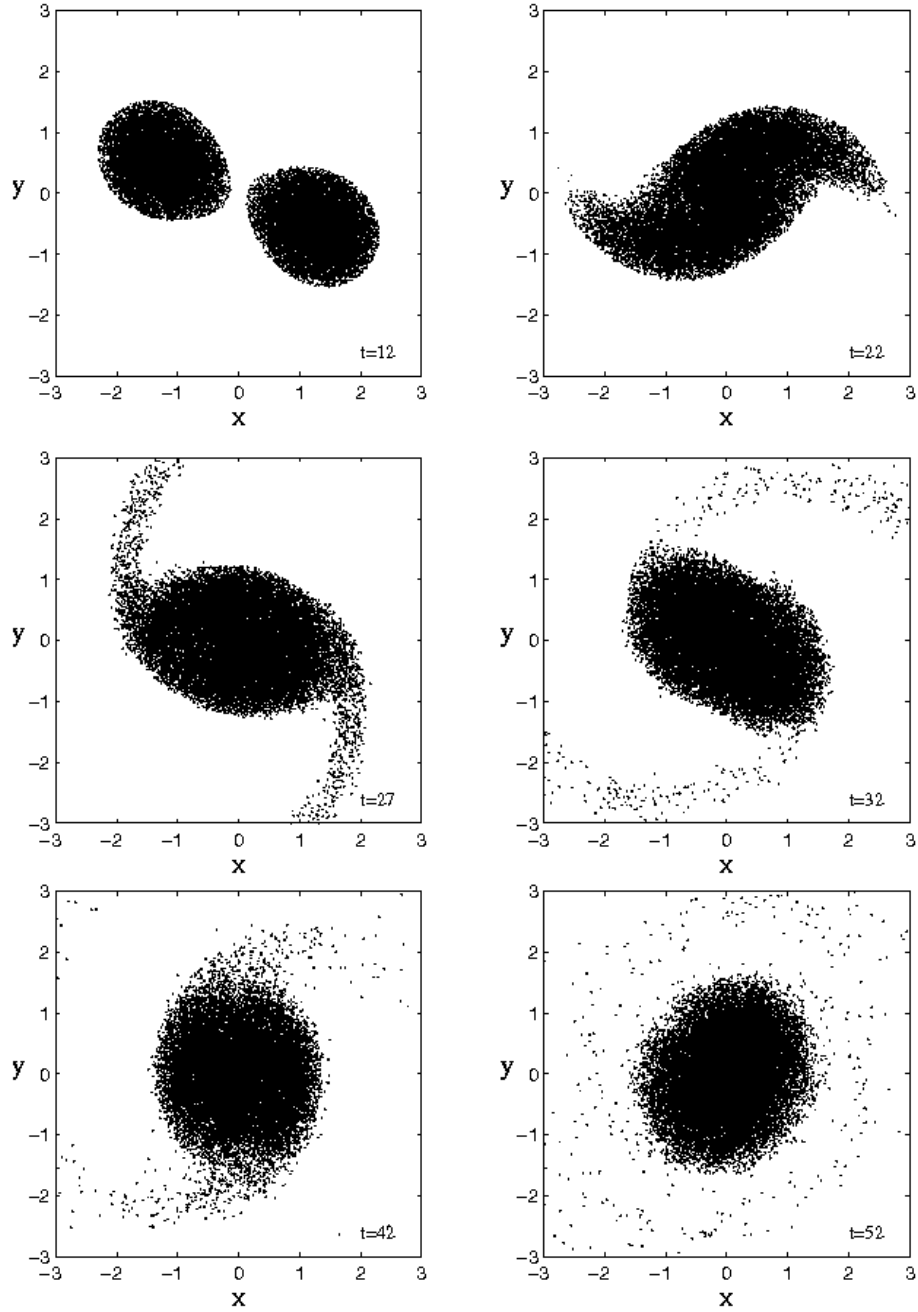}
\caption{Evolution of the system in the PN run.  Conventions are as
in Fig.~\protect\ref{fig:xyplotn}.  We see in the initial frame that
the long axes of the NS are misaligned before contact.  
Note also that the mass-shedding is suppressed compared to the N case.}
\label{fig:xyplotp}  
\end{figure}
\newpage
\begin{figure}
\centering \leavevmode \epsfxsize=7in \epsfbox{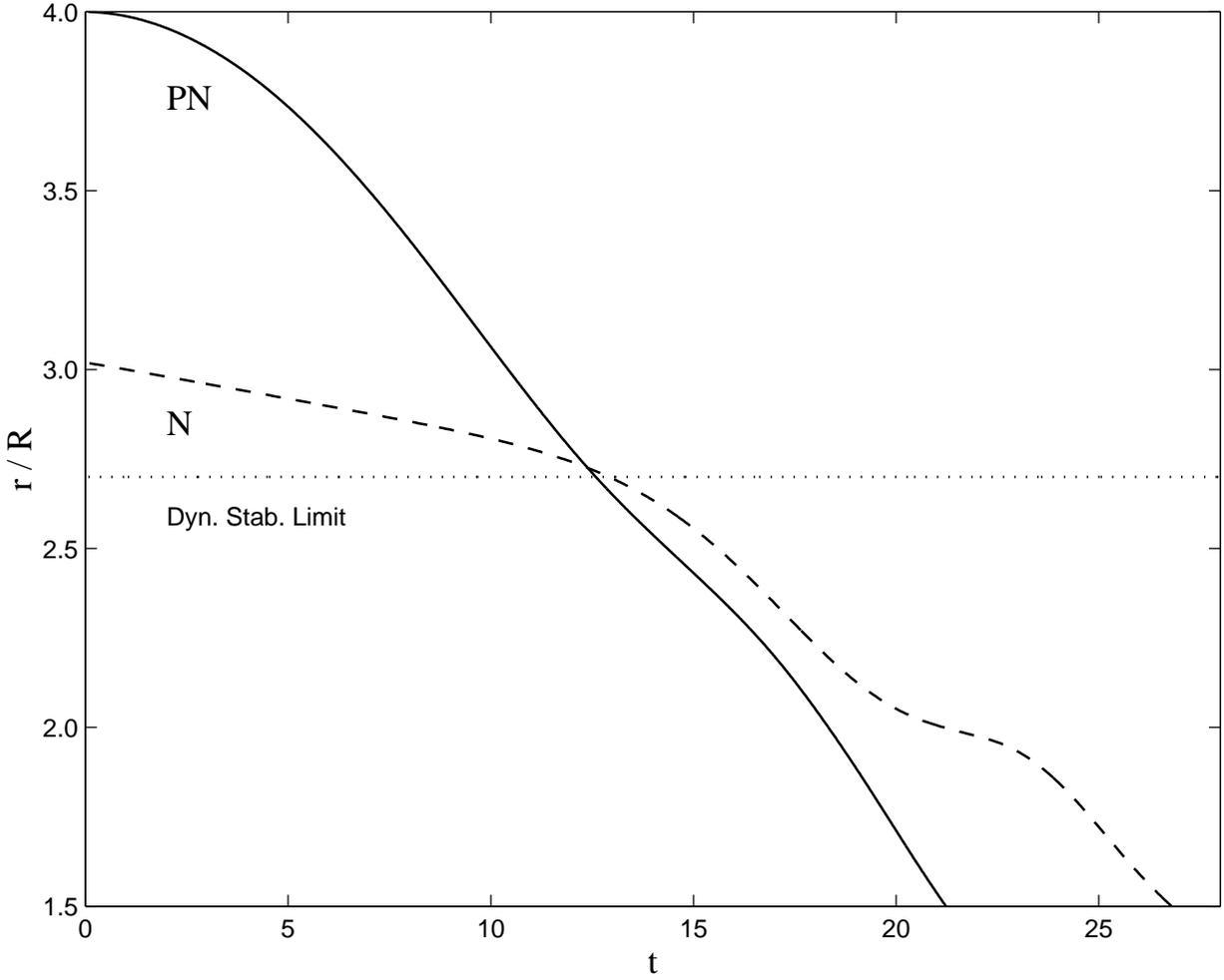}
\caption{Evolution of the 
binary center of mass separation during the inspiral phase
for the two calculations.  
The solid line is for the PN run, the dashed line for the N  run.  
The horizontal line
represents the dynamical stability limit for a Newtonian, equilibrium
binary, at $r\simeq
2.7\,R$. It appears as a break in the inspiral rate of the Newtonian binary,
whereas the PN binary inspiral becomes dynamical at significantly greater
separation.} 
\label{fig:sepplots}
\end{figure}
\newpage
\begin{figure}
\centering \leavevmode \epsfxsize=7in \epsfbox{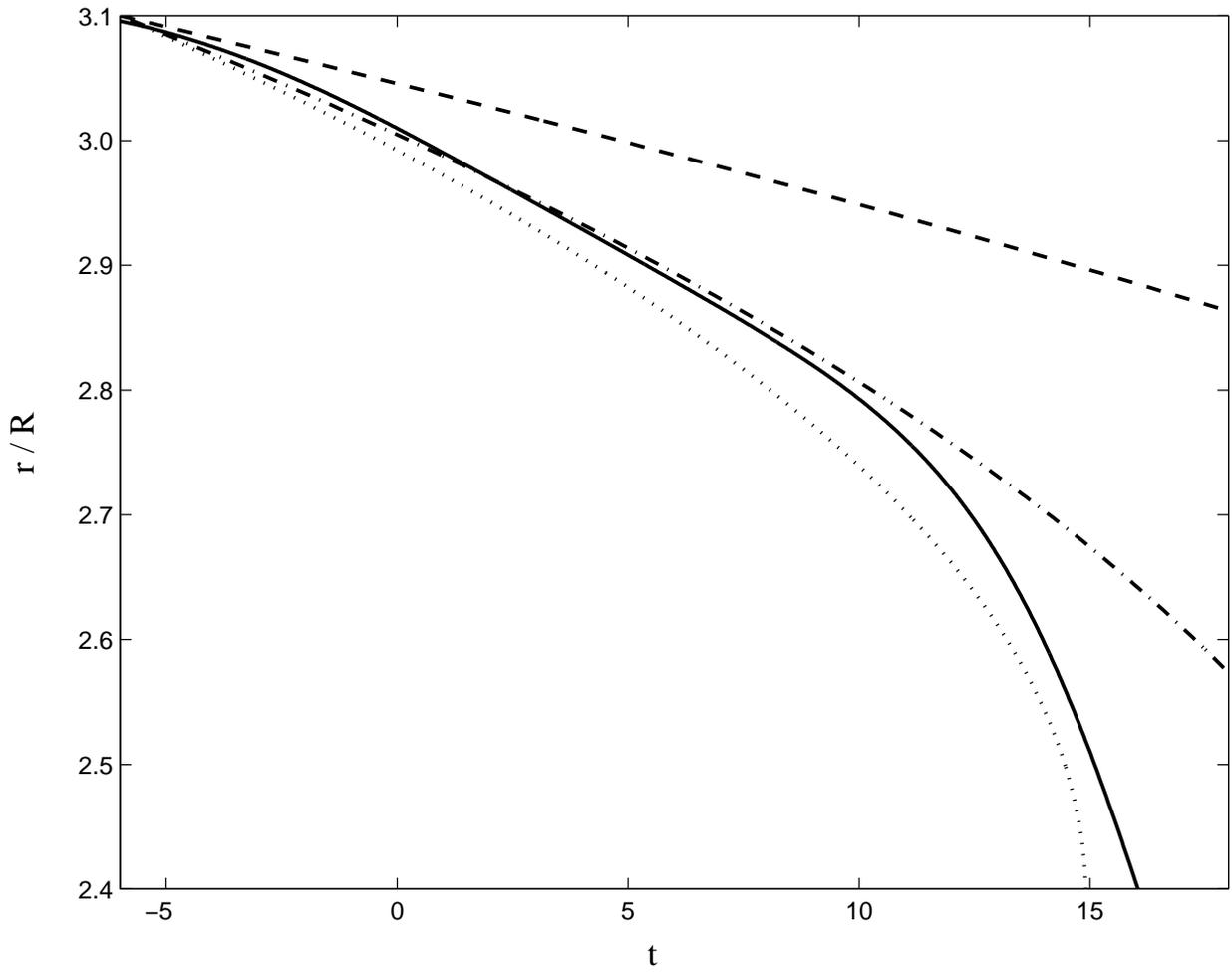}
\caption{Same as Fig.~\protect\ref{fig:sepplots}, but focusing on the
early inspiral of the Newtonian binary.  
The solid line is the result from the SPH calculation (N run).
The dashed line shows the point-mass
approximation, the dash-dotted and dotted lines the approximations for two
spheres and two ellipsoids, respectively.  See text for details.
The point mass approximation clearly fails
when tidal interactions become significant, but note the excellent
agreement with semi-analytic results for extended stars before the
ISCO is encountered.}
\label{fig:sepn}  
\end{figure}
\newpage
\begin{figure}
\centering \leavevmode \epsfxsize=7in \epsfbox{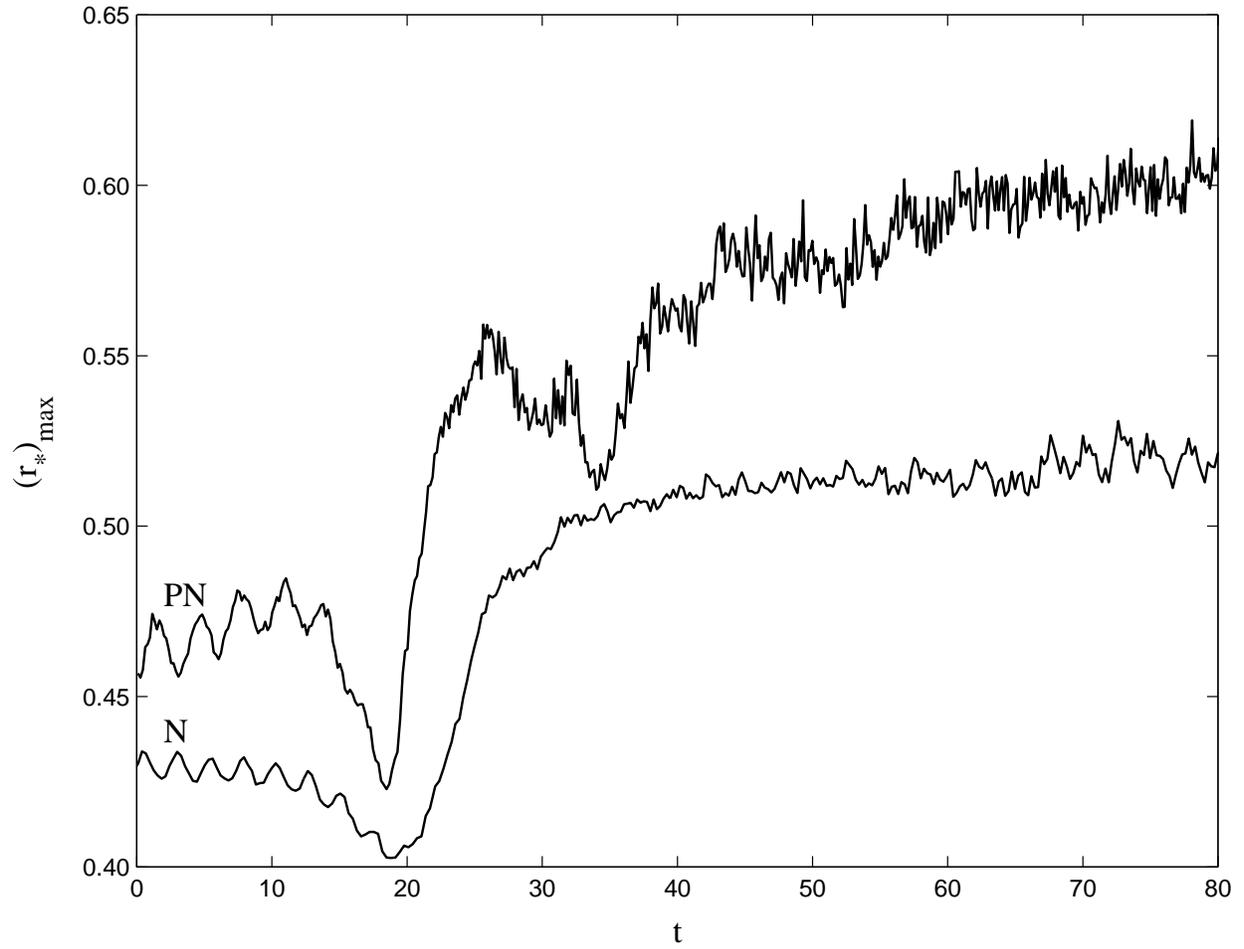}
\caption{Evolution of the maximum density in the two coalescence
calculations.  The upper curve is for the PN run, the lower curve for
the N run. 
The sharp decline in density at $t\simeq 15$ occurs as the two NS are
tidally disrupted, followed by a larger increase as
they coalesce.}
\label{fig:rhoplots}  
\end{figure}
\newpage
\begin{figure}
\centering \leavevmode \epsfysize=8in \epsfbox{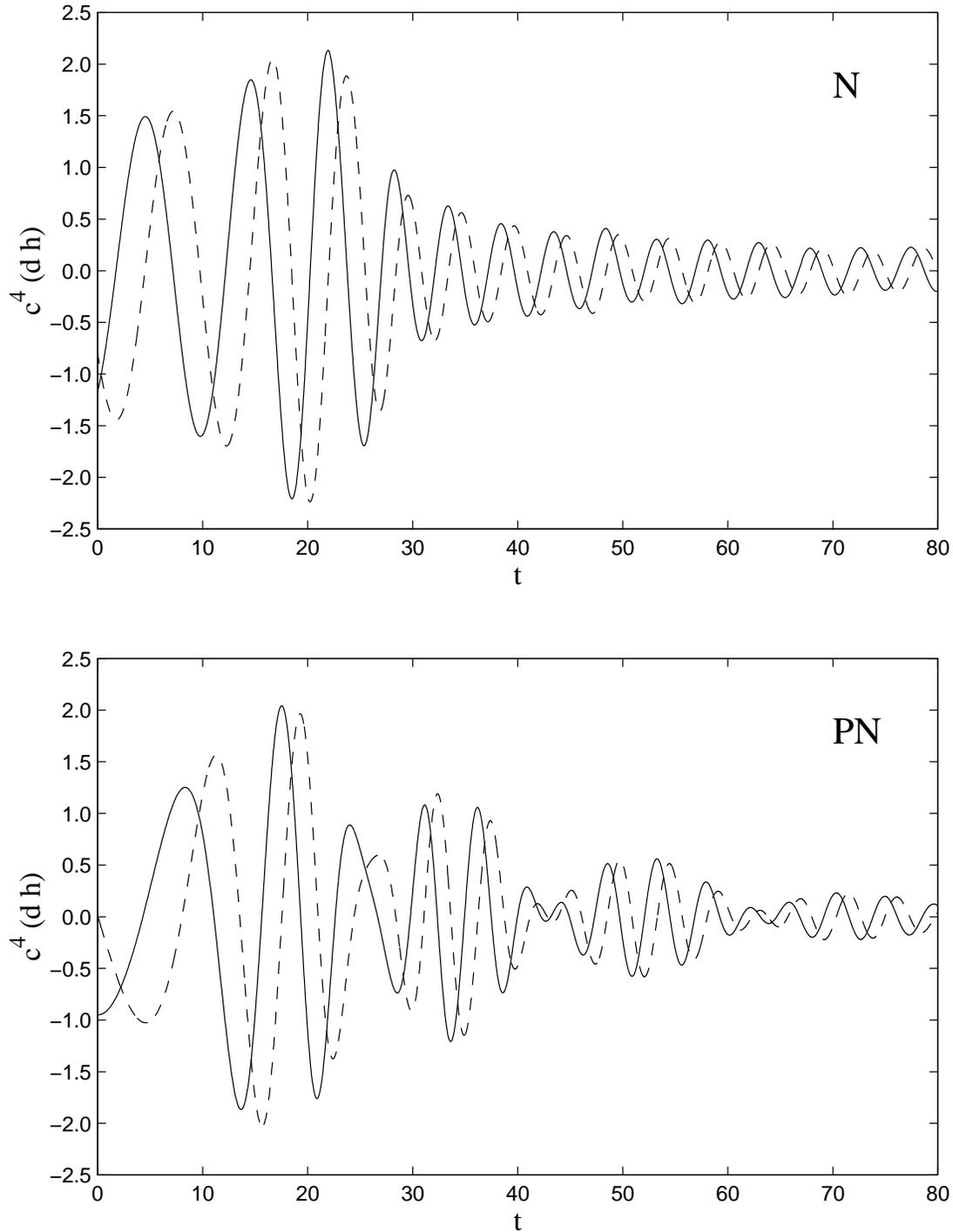}
\caption{Gravity wave signatures for the two coalescence runs.  
The waveforms are calculated for an observer at a distance $d$ along
the rotation axis.  The solid
line shows the $h_+$ polarization, the dashed line  the $h_{\times}$ 
polarization (see Eqs.~\protect\ref{hplus},\protect\ref{hcross}).
At late times in the N run the waveforms show a simple, 
exponentially damped oscillation, whereas in the PN run
an additional large-amplitude modulation is apparent.}
\label{fig:gwplots}  
\end{figure}
\newpage
\begin{figure}
\centering \leavevmode \epsfxsize=7in \epsfbox{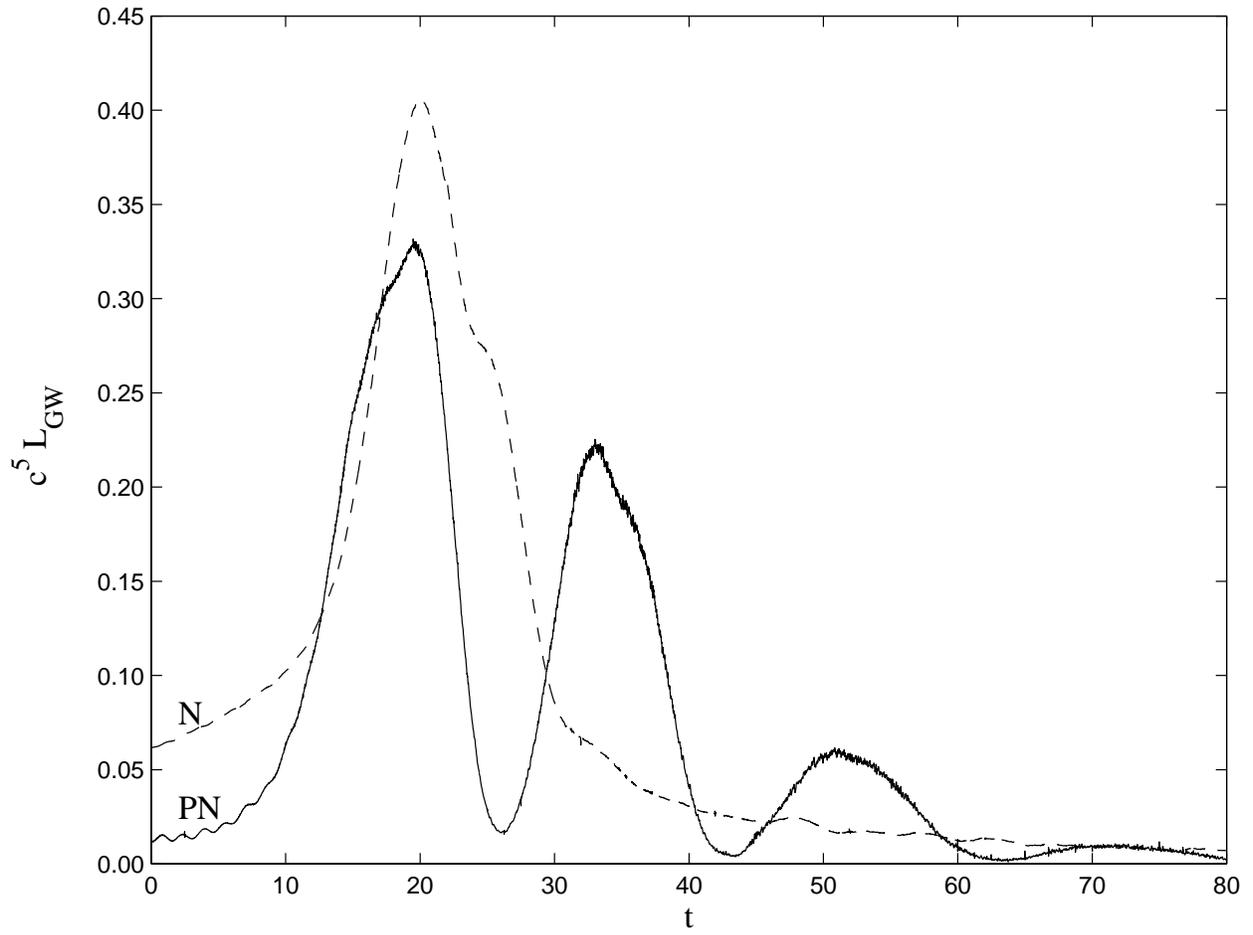}
\caption{Gravity wave luminosity for the two coalescence runs (see
Eq.~\protect\ref{gwlum}).  The solid
line is for the PN run, the dashed line for the N run.  The peak luminosity
in the PN run is smaller than that of the N run, but 
secondary peaks occur at $t\simeq 35$, $50$, and $70$.}
\label{fig:gwlum}  
\end{figure}
\newpage
\begin{figure}
\centering \leavevmode \epsfxsize=7in \epsfbox{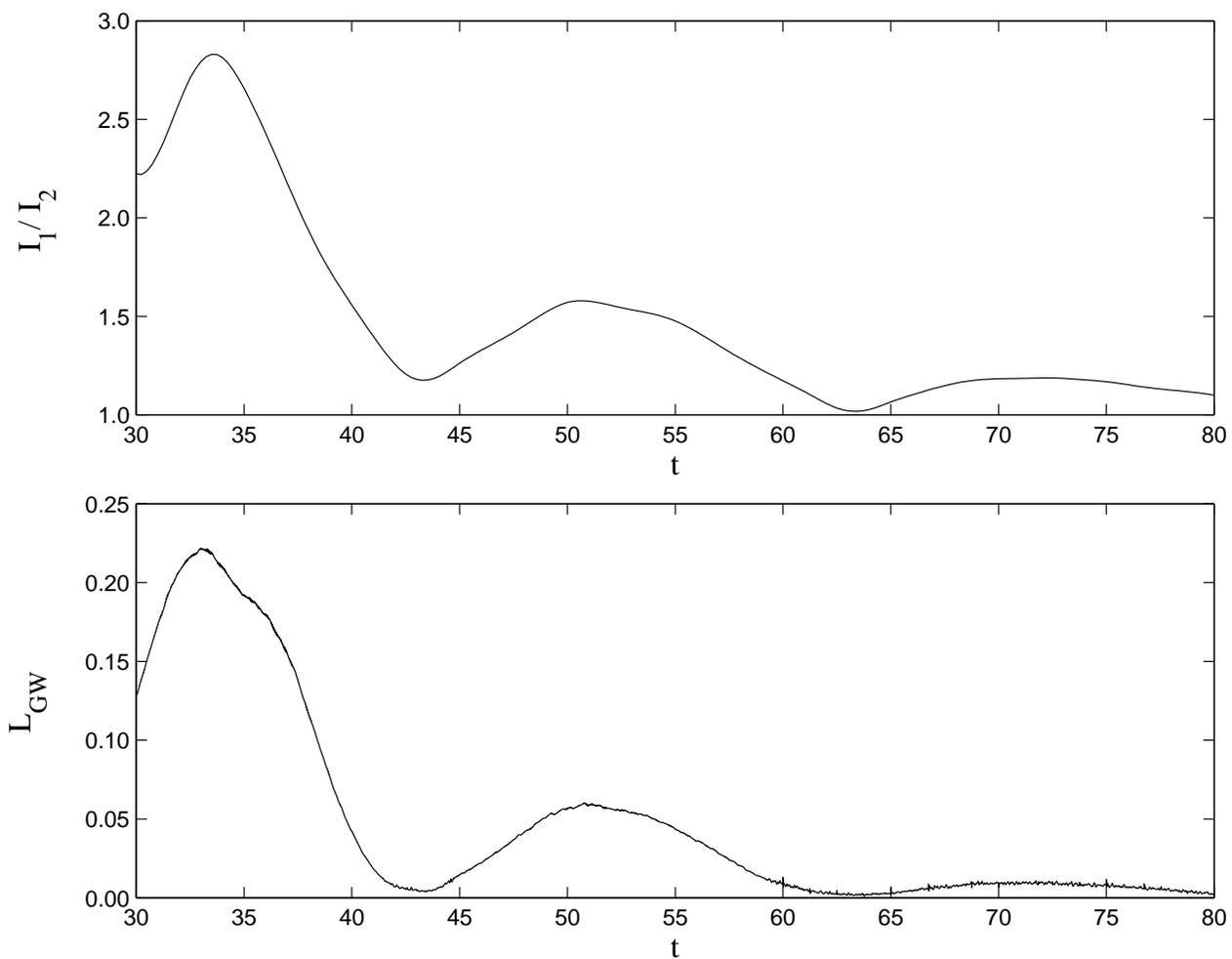}
\caption{The ratio of the principal moments of inertia in the
equatorial plane for the PN merger remnant, compared to the gravity
wave luminosity at late times.  
The times of maximum elongation correspond to maxima in the
gravity wave luminosity, and to decreases in the maximum density in
Fig.~\protect\ref{fig:rhoplots} at $t\simeq35$ and $t\simeq50$ 
(and less clearly at $t\simeq70$).}
\label{fig:mom}
\end{figure}
\newpage
\begin{figure}
\centering \leavevmode \epsfysize=8in \epsfbox{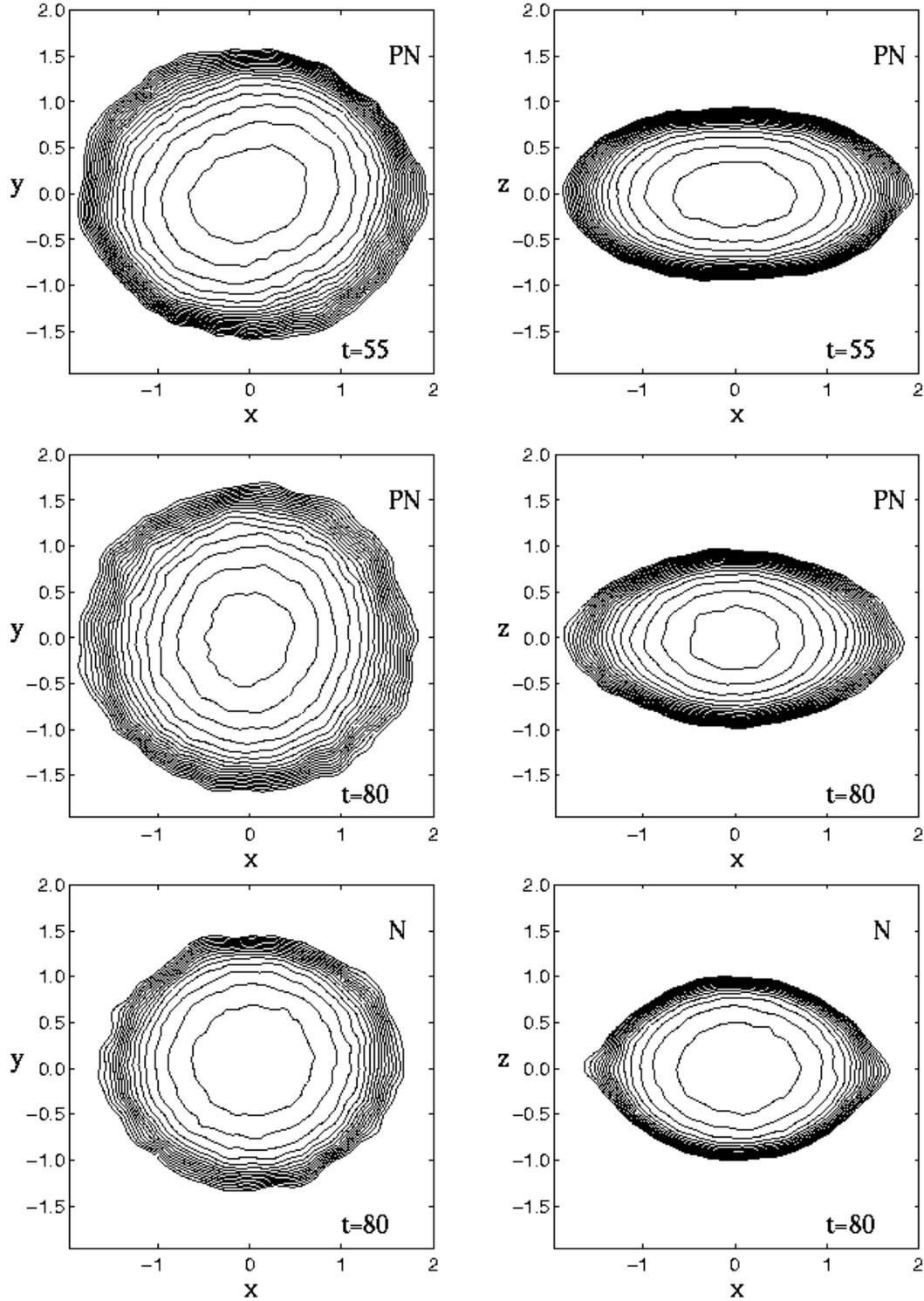}
\caption{Density contours of the merger remnants.  
The top frames show the PN remnant at $t=55$,
the middle ones show the same remnant at $t=80$, and the lower ones
show the N remnant at $t=80$. The left frames show a cut
through the equatorial plane, the right frames through the
vertical plane (containing the rotation axis).
Contours are logarithmic, ten per decade, starting from the maximum
density of $(r_*)_{max}=0.567$ for the PN run at $t=55$,
$(r_*)_{max}=0.608$ for the PN run at $t=80$, and $(r_*)_{max}=0.518$
for the N run at $t=80$.
The axes have been rotated to fall along the principal axes of the
remnant.  Note the cusp-like shape of the contours near the equator in
the vertical plane, indicating maximal rotation.}
\label{fig:finalcontour}
\end{figure}
\newpage
\begin{figure}
\centering \leavevmode \epsfxsize=7in \epsfbox{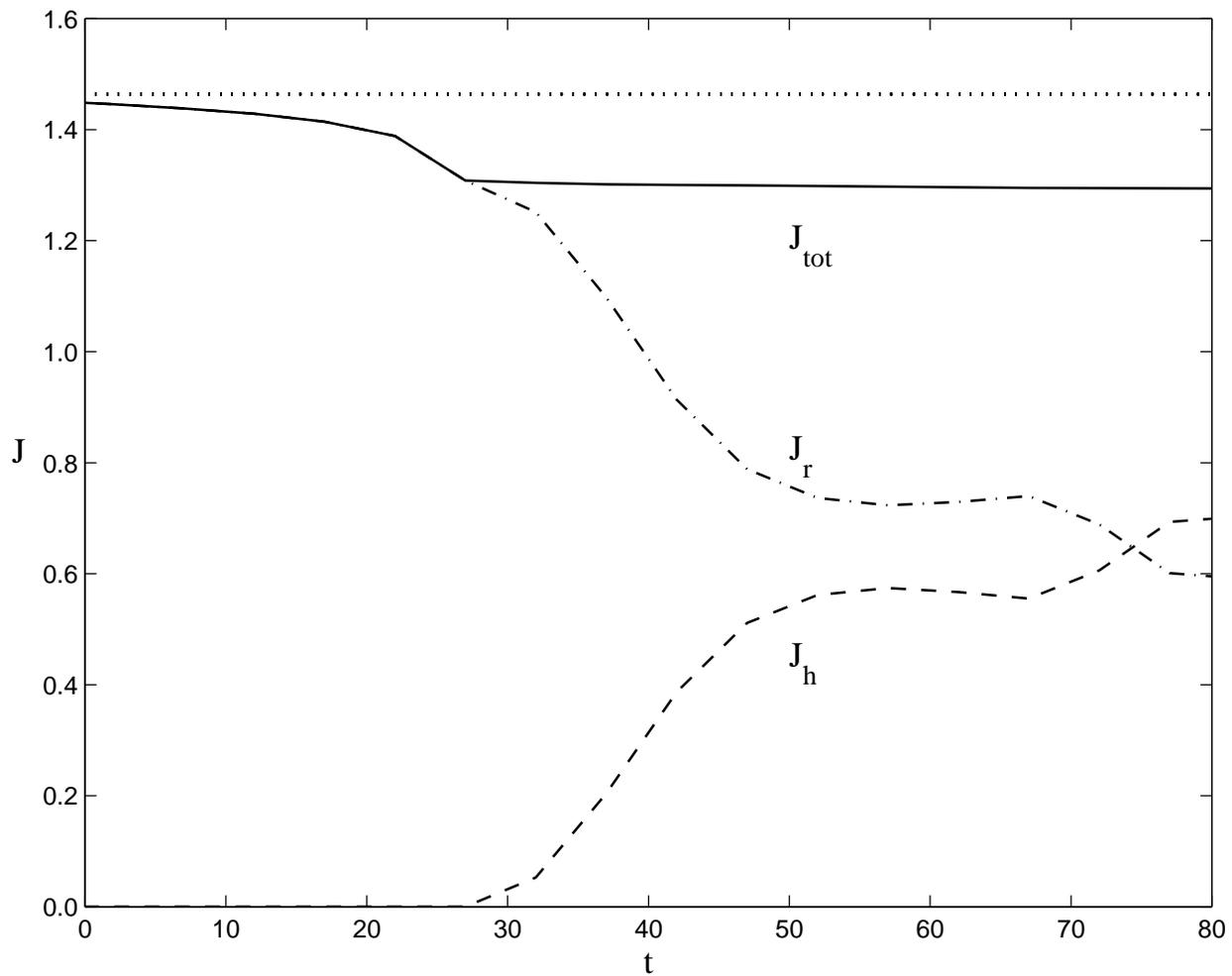}
\caption{Evolution of the angular momentum in various components in
the N run.  Here $J_{tot}$ is the total angular momentum in the
system, $J_r$ is for the inner remnant (defined by the condition
$r_*>0.005$, which includes the entire binary initially, but only the
inner remnant at later times), and $J_h$ is for the outer halo (so
that $J_{tot}=J_r+J_h$).  The dotted line shows the initial angular
momentum of the system.}
\label{fig:jm3n}
\end{figure}
\newpage
\begin{figure}
\centering \leavevmode \epsfxsize=7in \epsfbox{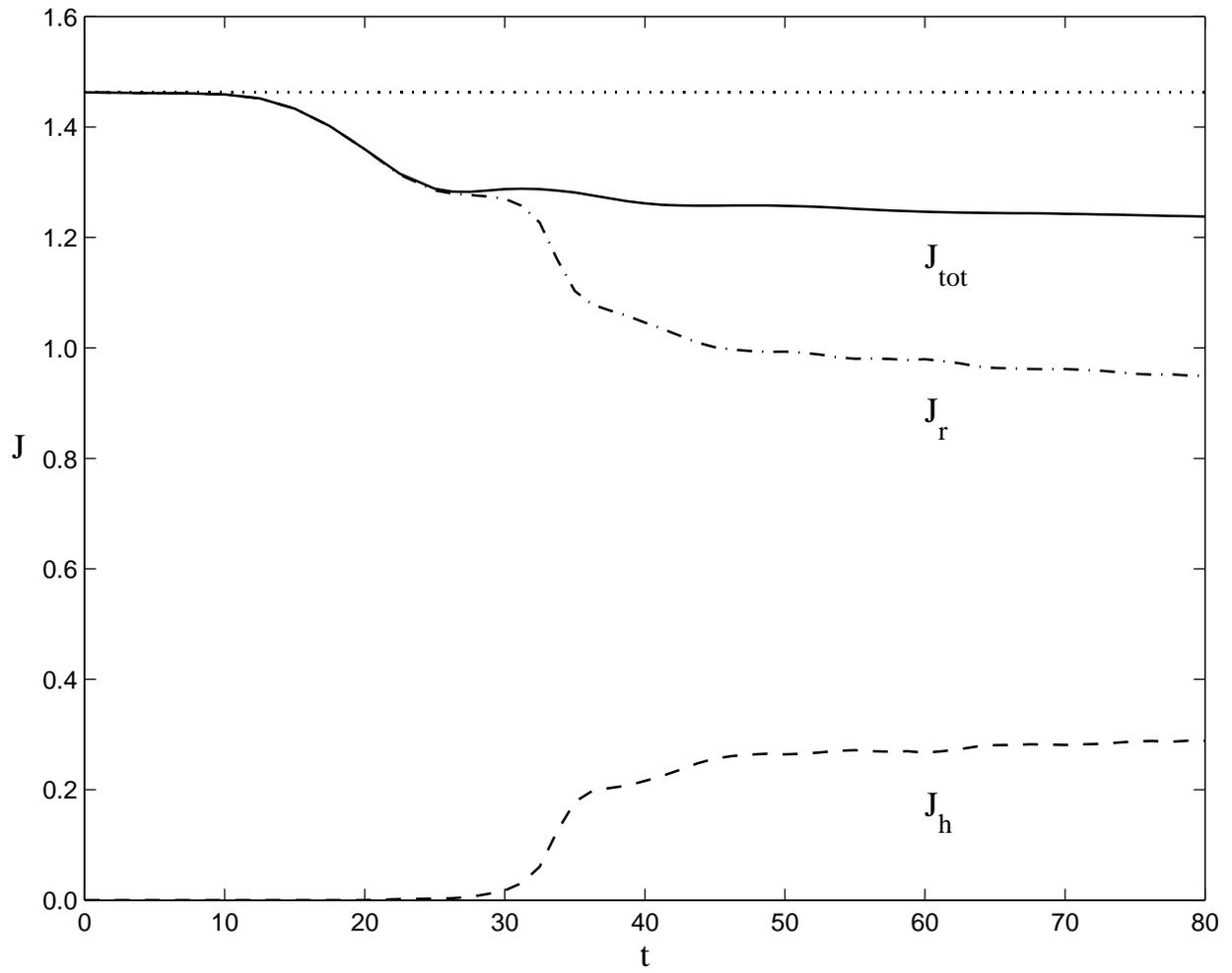}
\caption{Evolution of the angular momentum in various components in
the PN run.  Conventions are as in Fig.~\protect\ref{fig:jm3n}.}
\label{fig:jm3p}
\end{figure}
\newpage
\begin{figure}
\centering \leavevmode \epsfxsize=7in \epsfbox{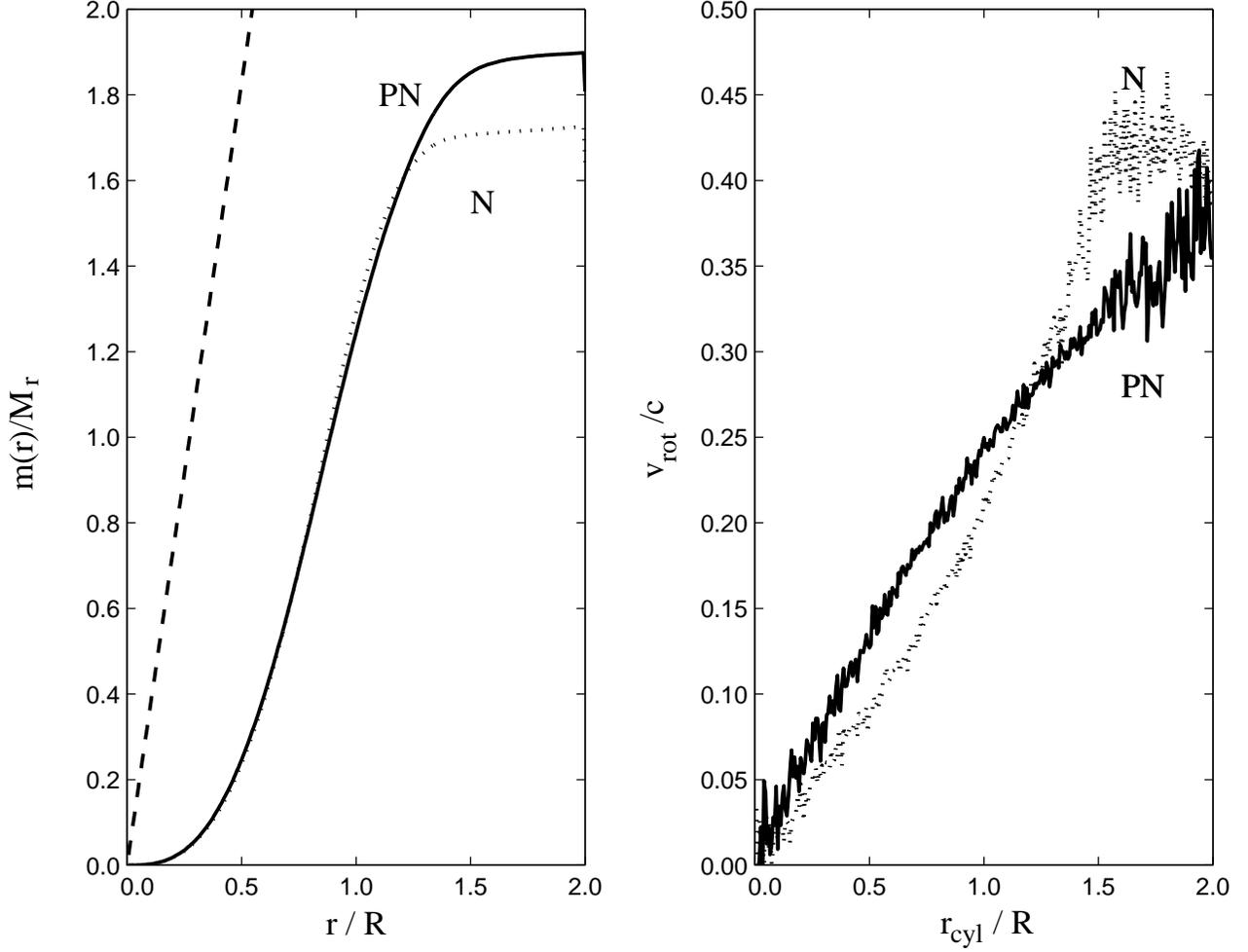}
\caption{Enclosed rest mass and radially averaged rotational velocity  
profiles of the final merger remnant at t=80 for the two runs. 
Here, $r_{cyl}$ is the distance from the rotation axis, while $r$ is
the radius from center.   In both
 plots, the solid line is for the PN run, the dotted line for the
 N run.  The dashed line shows the radius
 for a Kerr black hole with $a=0.7$ (the value we find for the PN run at t=80).
For the rotational profile, we show only the data
 for $-0.1<z<0.1$, all other horizontal cuts yielding similar
 profiles extending to smaller radii.}  
\label{fig:finalstate}  
\end{figure}
\newpage
\begin{figure}
\centering \leavevmode \epsfxsize=7in \epsfbox{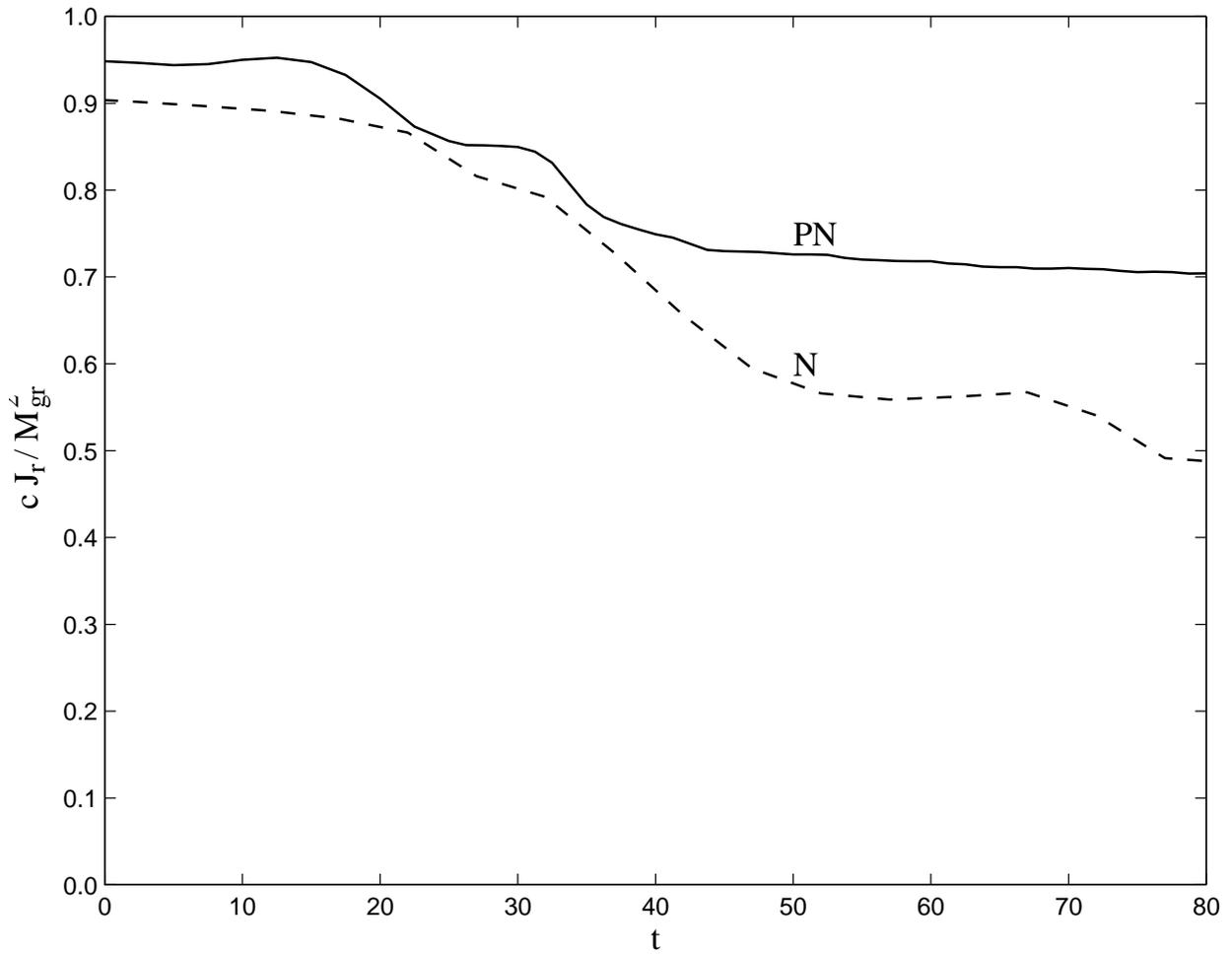}
\caption{The evolution of the Kerr parameter
$a_r\equiv cJ_r/M_{gr}^2$, for the inner remnant in the PN run (solid
line) and in the N run (dashed line).  
At no time do we have $a_r>1$.
The inner remnant (or core) is defined by the same density cut
as in Fig.~\protect\ref{fig:jm3n}.}
\label{fig:jm2}
\end{figure}
\newpage
\begin{figure}
\centering \leavevmode \epsfxsize=7in \epsfbox{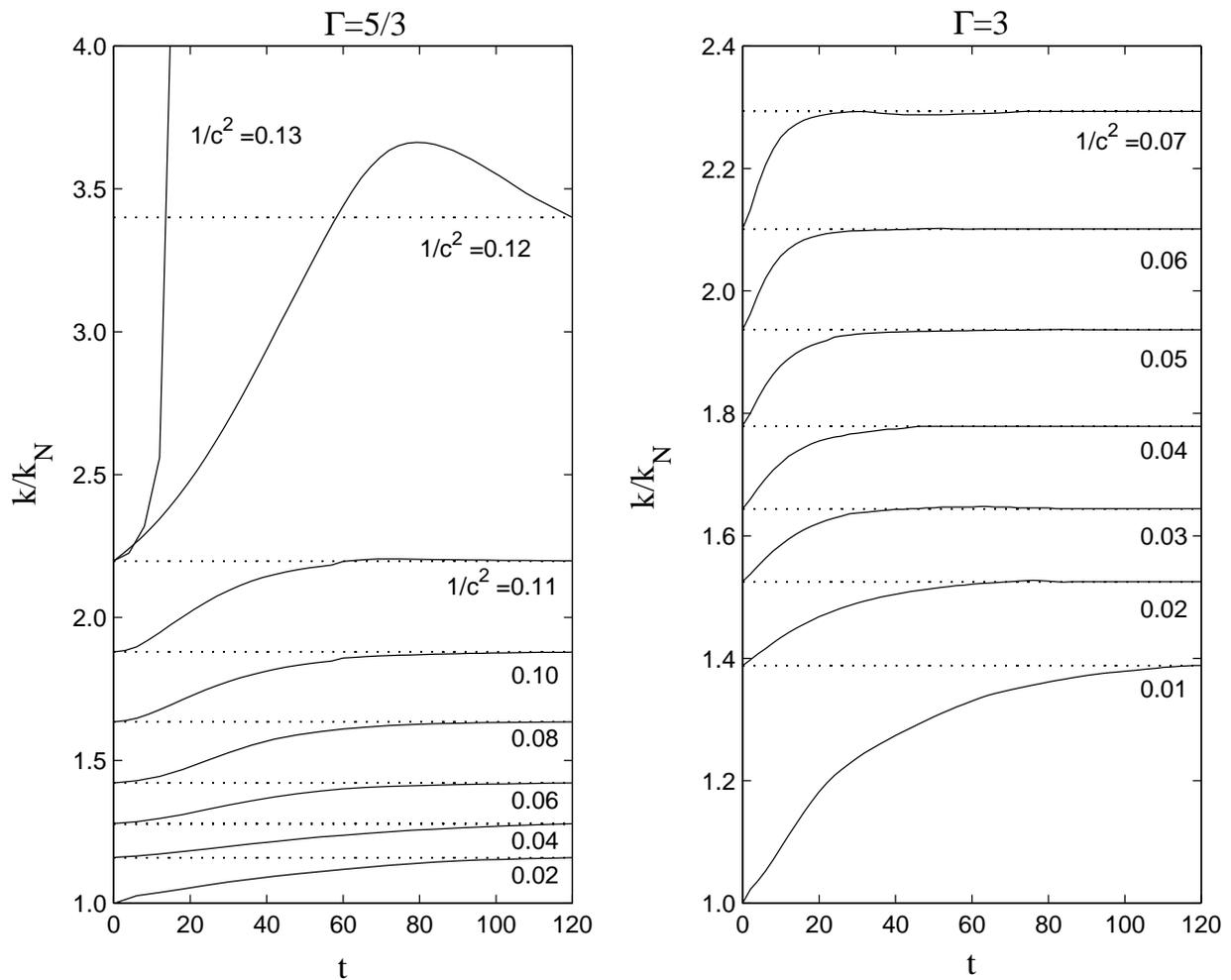}
\caption{Results of SPH relaxation calculations for single stars.
The ratio of the PN specific entropy
$k$ to the Newtonian value $k_N$, is shown for both $\Gamma=5/3$ and
$\Gamma=3$, computed for sequences of increasing compactness $1/c^2$.  
The dotted lines give the final value for each case,
which was used as the initial value for the next relaxation.  For
$\Gamma=5/3$, we see that for $1/c^2>0.12$, $k/k_N$ increases without
bounds, indicating instability.  For $\Gamma \protect =3$ and $1/c^2
\protect\gtrsim 
 0.07$, the 1PN approximation breaks down.}
\label{fig:plotak}
\end{figure}
\newpage
\begin{figure}
\centering \leavevmode \epsfxsize=7in \epsfbox{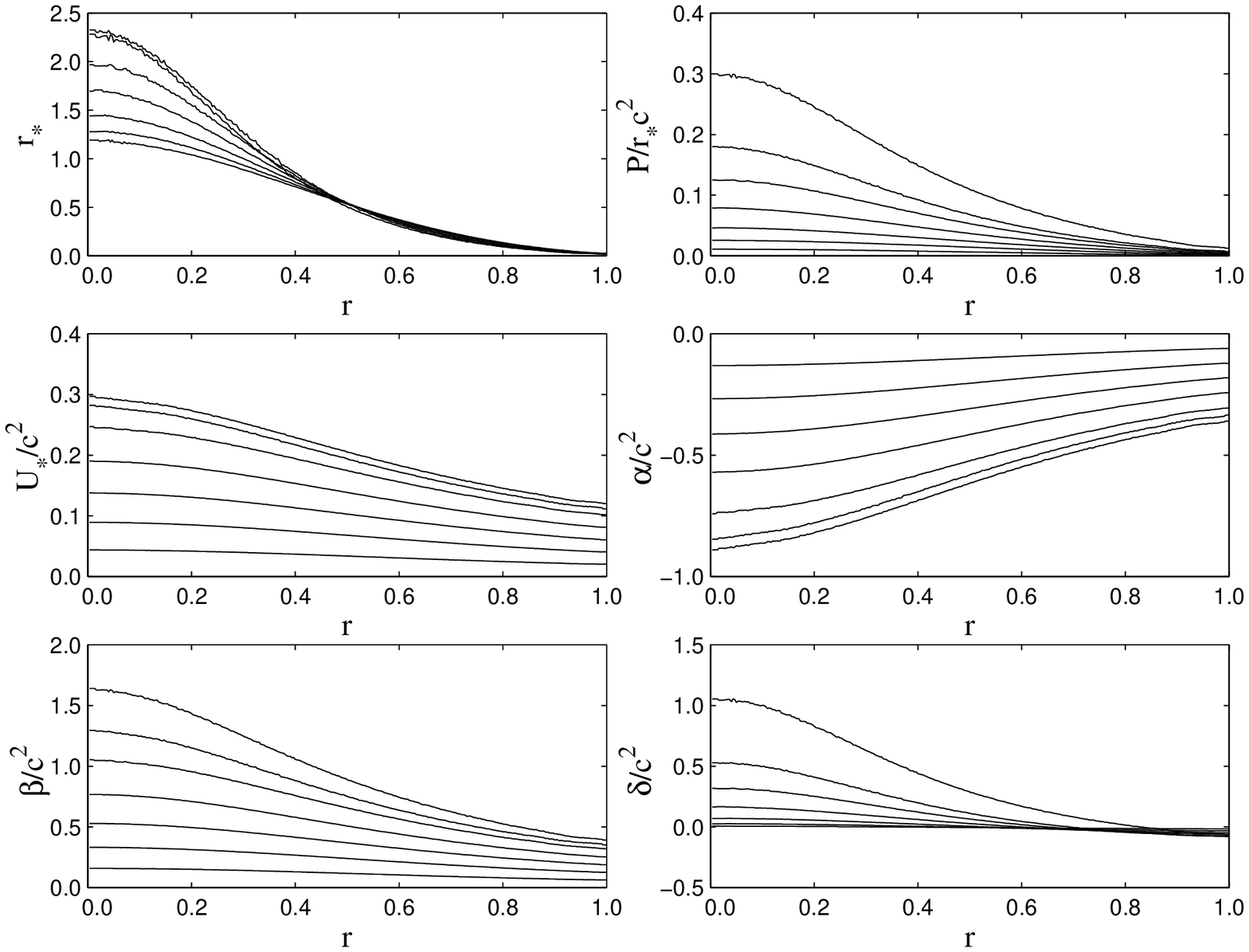}
\caption{Radial profiles for the $\Gamma=5/3$ single star models.
The various lines correspond, in monotonic fashion, to the stable
configurations indicated by dotted lines in the left panel of 
Fig.~\protect\ref{fig:plotak}.}
\label{fig:single}  
\end{figure}
\newpage
\begin{figure}
\centering \leavevmode \epsfxsize=7in \epsfbox{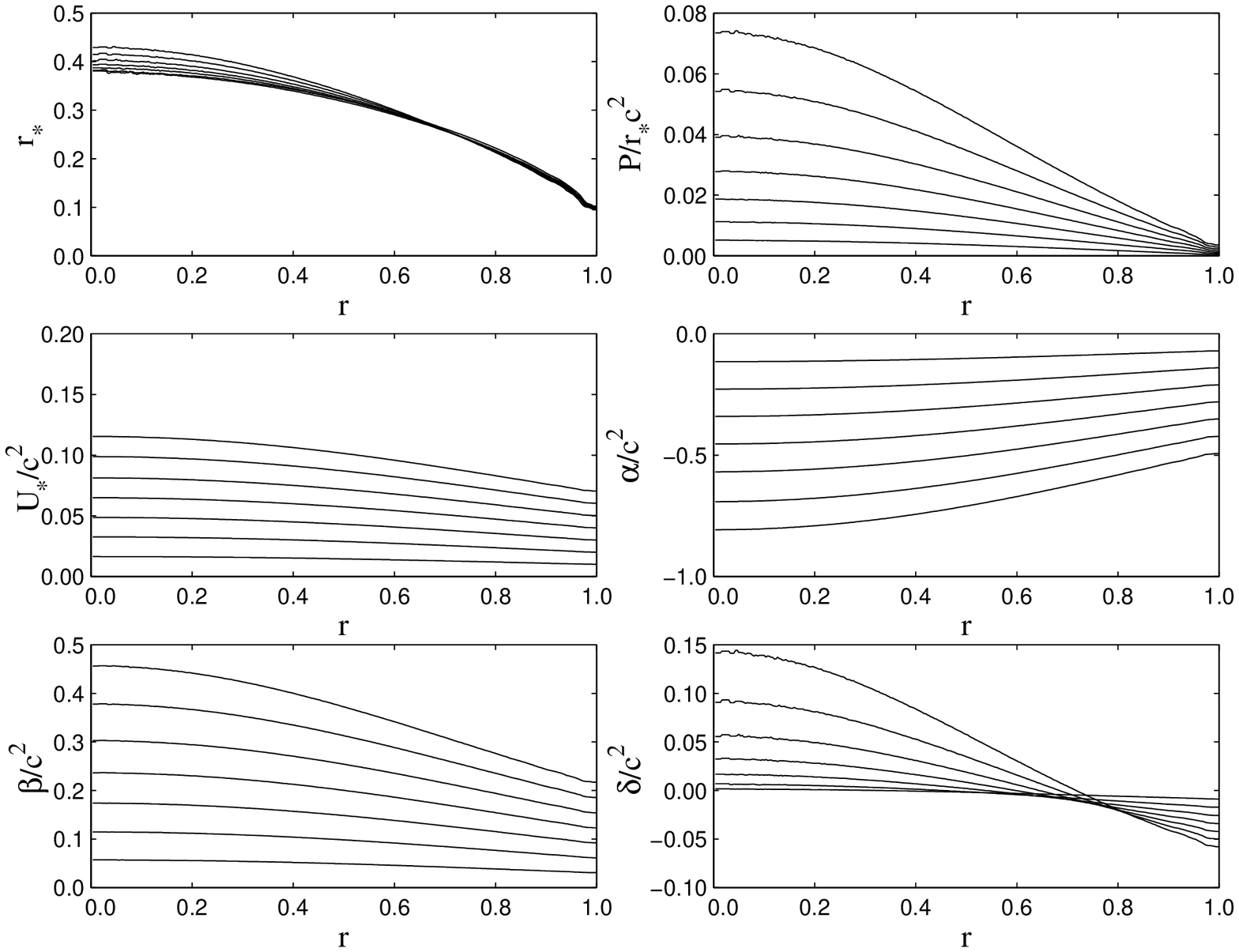}
\caption{Radial profiles for the $\Gamma=3$ single star models. 
The lines correspond, in monotonic fashion, to the stable 
configurations indicated by dotted lines in the right panel of 
Fig.~\protect\ref{fig:plotak}.}
\label{fig:single2}
\end{figure}
\newpage
\begin{figure}
\centering \leavevmode \epsfxsize=7in \epsfbox{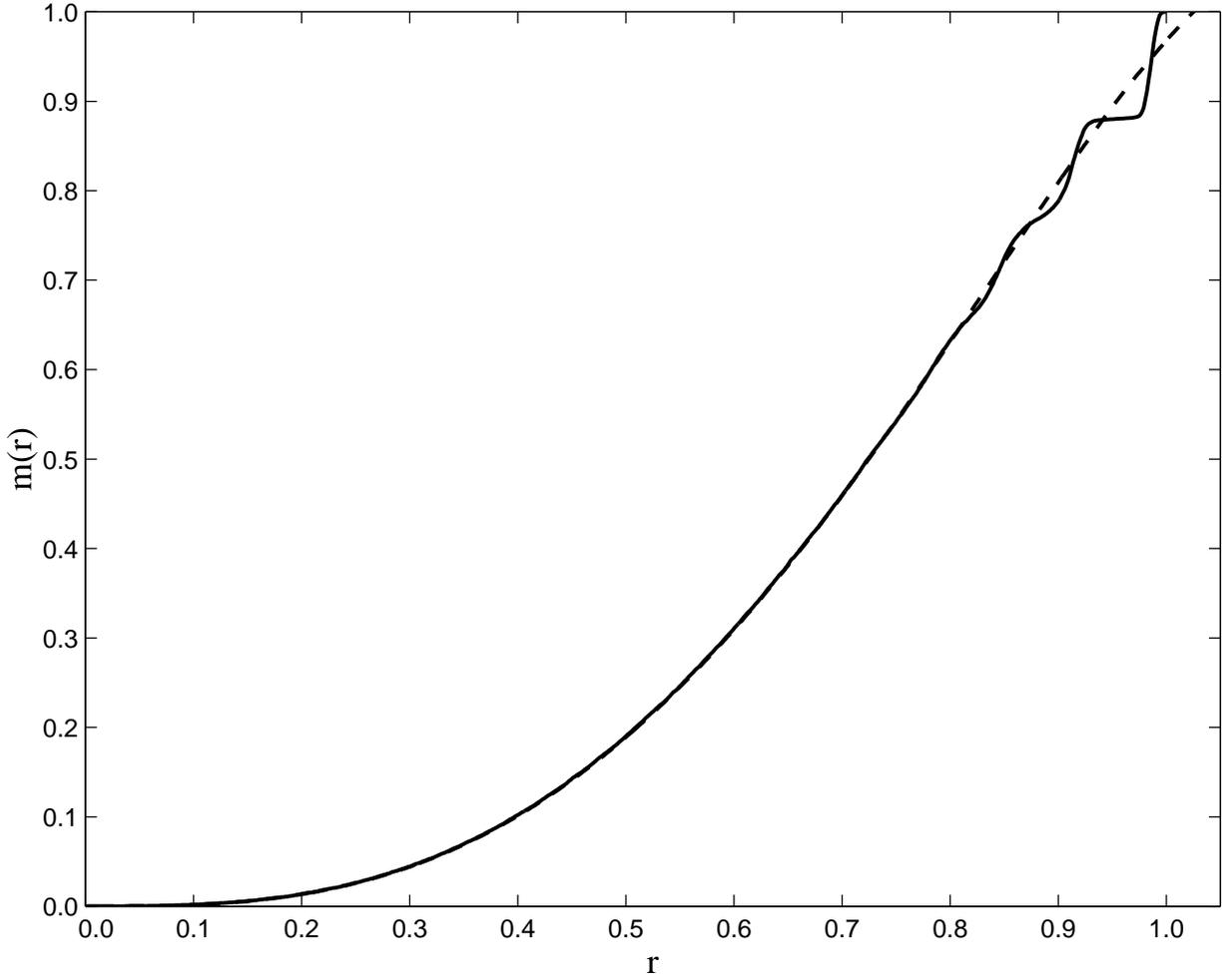}
\caption{Mass profile of the $\Gamma=3$ single NS model with
$1/c^2=0.05$, used in the PN run (solid), compared to a solution of
Eqs.~(\protect\ref{singlestar}--\protect\ref{singlestar2}) 
obtained by a direct numerical integration (dashed). 
The agreement is excellent, except near the outer surface of the
star, where the SPH density profile is more poorly defined.}
\label{fig:masrk}
\end{figure}
\newpage
\begin{table}
\caption{Properties of the Merger Remnants.
Units are such that $G=M=R=1$, where $M$ and $R$ are the mass and
radius of a single, spherical NS.  Here, $M_r$ is the rest mass of the
remnant, $M_{gr}$ is its gravitational mass, $J_r$ is its total angular
momentum, $\Omega_c$ and $\Omega_{eq}$ are the angular rotation
velocities at the center and at the equator, and the $a_i$'s and
$I_i$'s are the principle axes and moments of inertia.}
\label{table:final}

\begin{tabular}{cddd}
& N ($t=80$) & PN($t=55$) & PN($t=80$)\\
\tableline
$M_r$ & 1.73 & 1.90 & 1.89  \\
$M_{gr}$ & N/A & 1.85 & 1.84 \\
$J_r$ & 0.56 & 0.98 & 0.95 \\
$cJ_r/M_{gr}^2$ & 0.47 & 0.72 & 0.70 \\
\tableline
$\Omega_c$ & 0.45 & 0.80 & 0.75 \\
$\Omega_{eq}$ & 0.67 & 0.47 & 0.48 \\
\tableline
$a_1$ & 1.65 & 1.90 & 1.82 \\
$a_2$ & 1.35 & 1.58 & 1.53 \\
$a_3$ & 0.95 & 0.92 & 0.93 \\
\tableline
$a_2/a_1$ & 0.82 & 0.83 & 0.84 \\
$a_3/a_1$ & 0.58 & 0.48 & 0.51 \\
\tableline
$I_1$ & 0.575 & 0.861 & 0.741 \\
$I_2$ & 0.477 & 0.583 & 0.674 \\
$I_3$ & 0.247 & 0.228 & 0.236 \\
\tableline
$I_2/I_1$ & 0.829 & 0.677 & 0.909 \\
$I_3/I_1$ & 0.429 & 0.274 & 0.319 \\
\end{tabular}
\end{table}

\begin{table}
\caption{Parameters for Single Star Models. 
For each model, we list the
compactness parameter $1/c^2$, the ratio of the PN
specific entropy $k$ to the Newtonian value $k_N$, and the central
values of density $r_*$, and the dimensionless ratios $P/r_*c^2$ and
$U_*/c^2$.}
\label{table:stuffvsc} 
\begin{tabular}{ccccc}
$1/c^2$&$k/k_N$&$(r_{*})_c$&$(P/r_*c^2)_c$&$(U_*/c^2)_c$\\
\tableline
 & & $\Gamma=5/3$ & &\\
\tableline
0.02 & 1.177 & 1.201 & 0.0111 & 0.0438 \\
0.04 & 1.281 & 1.292 & 0.0257 & 0.0893 \\
0.06 & 1.421 & 1.452 & 0.0464 & 0.1377 \\
0.08 & 1.634 & 1.708 & 0.0792 & 0.1902 \\
0.10 & 1.879 & 1.976 & 0.1255 & 0.2470 \\
0.11 & 2.198 & 2.336 & 0.1806 & 0.2820 \\
0.12 & 3.400 & 2.295 & 0.3011 & 0.2968 \\
\tableline
 & & $\Gamma=3$ & &\\
\tableline
0.01 & 1.403 & 0.3822 & 0.0052 & 0.0165 \\
0.02 & 1.553 & 0.3818 & 0.0114 & 0.0326 \\
0.03 & 1.649 & 0.3882 & 0.0187 & 0.0486 \\
0.04 & 1.780 & 0.3948 & 0.0280 & 0.0649 \\
0.05 & 1.918 & 0.4051 & 0.0397 & 0.0813 \\
0.06 & 2.084 & 0.4170 & 0.0549 & 0.0989 \\
0.07 & 2.262 & 0.4321 & 0.0746 & 0.1154 \\ 
\end{tabular}
\end{table}
\end{document}